\documentclass[useAMS,usenatbib]{mn2e}
\usepackage{graphicx}
\usepackage{hyperref}

\title[A stable configuration for the extrasolar system HD 181433]{Treating dynamical stability as an observable: a 5:2 MMR configuration for the extrasolar system HD 181433}
\author[Giammarco Campanella]{Giammarco Campanella$^{1}$\thanks{E-mail:
g.campanella@qmul.ac.uk}\\
$^{1}$Astronomy Unit, School of Mathematical Sciences, Queen Mary University of London, 327 Mile End Road, London, E1 4NS, United Kingdom}

\begin{document}

\date{Accepted 2011 August 1. Received 2011 July 17; in original form 2011 June 8}

\pagerange{\pageref{firstpage}--\pageref{lastpage}} \pubyear{2011}

\maketitle

\label{firstpage}

\begin{abstract}
The three-planet extrasolar system of HD 181433 has been detected with HARPS. The best-fit solution, announced by the discovery team, describes a highly unstable, self-disrupting configuration. In fact, a narrow observational window, only partially covering the longest orbital period, can lead to solutions representing unrealistic scenarios. Taking into account the \textit{dynamical stability} as an additional observable while interpreting the RV data, we can analyse the phase space in a neighbourhood of the statistically best-fit and derive dynamically stable configurations that reproduce the observed RV signal. Our Newtonian stable best-fit model is capable of surviving for at least 250 Myrs. The two giant companions are found to be locked in the 5:2 MMR as Jupiter and Saturn in the Solar System. This mechanism does not allow close encounters even in case of highly eccentric orbits. Moreover, planets \textit{c} and \textit{d} are located in regions spanned by many other strong low-order MMRs. We study the dynamics of some plausible scenarios and we illustrate the behaviours caused by secular apsidal resonances and mean motion resonances. Furthermore, we find a terrestrial planet in the habitable zone of HD 181433 can retain stability. Apart from filling an empty gap in the system, this body could offer a harbour for life indeed. Additional measurements are necessary in order to investigate this hypothesis and can confirm the predictions outlined in the paper.
\end{abstract}


\section{Introduction}

Nowadays, more than 50 multi-planet systems are known to harbour 2-6 planets\footnote{``The Extrasolar Planets Encyclopaedia'' \url{http://www.exoplanet.eu}}. The recent years have seen a proliferation of multiple-planet systems thanks to the increase in both instrument precisions and duration of several planet search programs. This has allowed the detection of longer periods planetary signatures, as well as planetary signatures with lower amplitude.

An increasing sample of found planets improves our knowledge of their distribution in the mass-period diagram and allows comparison with theoretical predictions (e.g. Wright et al. 2009). In fact, dynamical analysis of planetary systems can both precisely determine their orbital architectures and constrain their evolutionary histories providing a test bed for planetary formation and evolution theories. To make these investigations we need measured orbital parameters as accurate as possible. Unfortunately, the accuracy of such measurements are limited due to uncertainties and degeneracies inherent to the Radial Velocity (RV) discovery technique. Nevertheless, the RV method is the most efficient technique for detecting extrasolar planets with more than 90\% of all currently known planets being either detected or characterized using this method.

As the time baseline becomes larger, it is possible to distinguish a trend in the RV signals due to long-period outer companions that have not completed a single orbit yet. In this case, it can happen either that the profile of $\chi^2$ is very smooth, or that it does not have a well definite minimum, as a result the confidence levels may comprise large intervals of the parameters fitted.

The RV signal does not provide any direct information on the real masses and the orbital inclinations of the planets; even if we make the assumption the system is coplanar and seen edge-on, an N-planet configuration is described by some 5N+1 parameters. The \textit{a priori} unspecified number of planets, narrow observational windows, stellar jitter and weakly constrained orbital parameters, can lead to not unique solutions and (quite often) to best-fits representing unrealistic scenarios. In fact, these solutions can present well-constrained minimum masses but also a poorly constrained eccentricity for the outermost planet; i. e., in the statistically optimal best-fit solutions, the eccentricities can be large and can rapidly (on the timescale of thousands of years) bring to catastrophic collisional instabilities \citep{b9}.

According to the Copernican assumption that we are not observing the universe at a privileged time, the detection of a rapidly unstable system during a few years of RV observations is not likely, then a fit which corresponds to a quickly unstable configuration is also doubtful. Therefore, we can aim to put limits on the masses and orbital elements of the planets by investigating and finding the plausible and stable solutions. In this logic, the \textit{dynamical stability} is an additional observable that must be taken into consideration when interpreting the RV data. It turns into a discriminating element especially when the longest orbital period is only partially covered \citep{b18,b6}.

Often, a stability criterion is applied once the best-fit unstable solution is found, subsequently the orbital elements are tuned to get a stable configuration. However this approach does not necessarily provide stable fits that are simultaneously optimal in term of $\chi^2$ or RMS. Most of the times, with the term \textit{stable} we indicate a configuration which does not disrupt or change qualitatively during a period of time of the order of million years. The literature is plenty of studies which take into account stability when modelling the RV data. Here we just mention the work of: \citet{b24} reporting on the system stability, secular evolution and the extent of the resonant interactions for 5 dynamically active multi-planet systems; \citet{b26} which derive updated orbital parameters for a number of systems considering mutual interactions between planets; \citet{b6,b7,b9} which directly eliminate unstable solutions during the fitting procedure using GAMP; and \citet{b5} which give constrains on the inclination with respect to the line of sight for some of the planets in the GJ 876 system. We also should remind the study about the directly imaged system HR 8799, where the difficulties in finding regions of stable motion may indicate the system is undergoing a phase of planet-planet scattering \citep{b8}.

Thanks to powerful instrumentations like the HARPS spectrograph at La Silla Observatory, the Radial Velocity accuracy is increasing rapidly breaking the barrier of 1 m/s \citep{b19}. This has permitted the detection of weaker signals consenting the discovery of some of the lowest-mass planets identified such as: GJ 876 d \citep{b21}, HD 40307 b \citep{b14}, 61 Vir b \citep{b25} and GJ 581 e \citep{b15}.

The planetary system of HD 181433 has been discovered with HARPS. It has been reported to contain 3 planets: two Jupiter-class planets and a Super-Earth of 7.5 $m_\oplus$ \citep{b4}.

Inspired by the peculiar properties of the system, which includes two giant planets and one rocky planet all in high eccentric orbits, we aimed to study the past and future evolution of the system. Unfortunately, the published configuration is unstable\footnote{See \url{http://xsp.astro.washington.edu}}. The model in which the initial eccentricities of the planets are reduced by one sigma quickly leads to disruption too. The fate does not change when we assume, in addition, a mutual orbital inclination of $20\,^{\circ}$ between planets \textit{c} and \textit{d}.

These attempts evidence the necessity of doing an analysis from scratch in order to get a self-consistent solution compatible with the data. In this paper we examine the available RV data of the HD 181433 system \citep{b4} taking a more general approach, going beyond a formal fit of the Keplerian orbital elements. Even if the RV observations do not span a single period of the outermost planet, we can give reasonable constraints on the orbital elements of the poorly sampled third planet by studying the dynamics of the system. In the Keplerian fit this important information is completely omitted.

In many situations, the interactions between planets are negligible and can be ignored. So the RV signal is just a linear superposition of different Keplerian RV curves. On the other hand, planetary interactions and resonances can be important and must be taken into account performing numerical integrations. In the short time scale, these interactions can cause significant variations in the orbital parameters of the planets. As in our case, an ensemble of constant Keplerian orbital elements is not adequate to model the RV data and an \textit{N}-body Newtonian model should be applied.

However, it can also happen that good Newtonian fits to the data produce planetary orbital parameters that are stable for the period of observations but lead to disruption on timescales substantially shorter than the age of the planetary system. In this respect, long-term stability is an additional but necessary constraint that must be satisfied by multiplanet fits.

In Section \ref{RV}, we perform an independent analysis of the RV data for HD 181433. We probe the phase space of the orbital parameters looking for likely configurations stable for long timescales, say millions of years. We assume the motion is described by Newtonian interacting orbits. In Section \ref{longterm}, we present the dynamical study of the stable best-fit solution and we analyse the behaviour of other plausible stable configurations. In particular, we focus on the description of secular apsidal resonances (SARs) and mean motion resonances (MMRs). In Section \ref{concl}, we briefly summarize our findings, we discuss on the possibility of a terrestrial planet in the habitable zone and we make some predictions about what we may expect from further observations.

\section[]{RADIAL VELOCITY DATA ANALYSIS}
\label{RV}

\citet{b4} announced the detection of 3 planets around HD 181433, a K3IV star, considering 107 RV measurements which covered more than 4 years, from June 2003 to March 2008. The median uncertainty for the RV is 0.53 m/s with most of the data in the range 0.4-1.0 m/s. The peak-to-peak velocity variation is 48.12 m/s, while the velocity scatter around the mean RV in the measurements is 13.86 m/s.

The data do not completely cover a full period for the third planet. In fact, what is possible to spot it is just an additional long-term trend which is modeled by \citet{b4} as being produced by a planet of minimum mass $m\sin i = 0.54$ $m_{Jup}$ on a Keplerian orbit with a period of about 6 years and $e = 0.48$. This model is unstable on the order of just thousands of years. The instability arises due to close encounters between planet \textit{c} and planet \textit{d}.

\begin{figure*}
\includegraphics[width=0.5\textwidth]{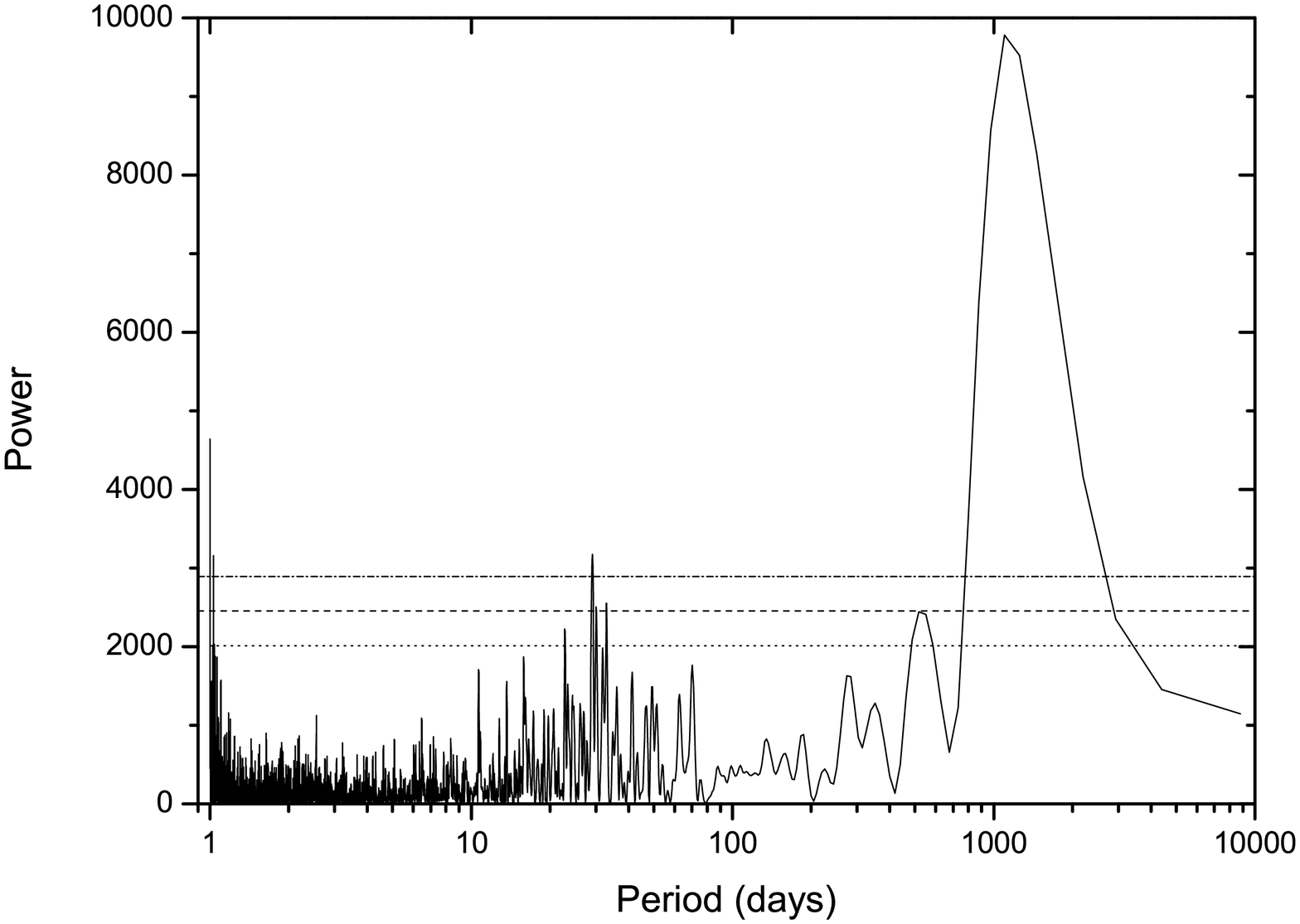}\includegraphics[width=0.5\textwidth]{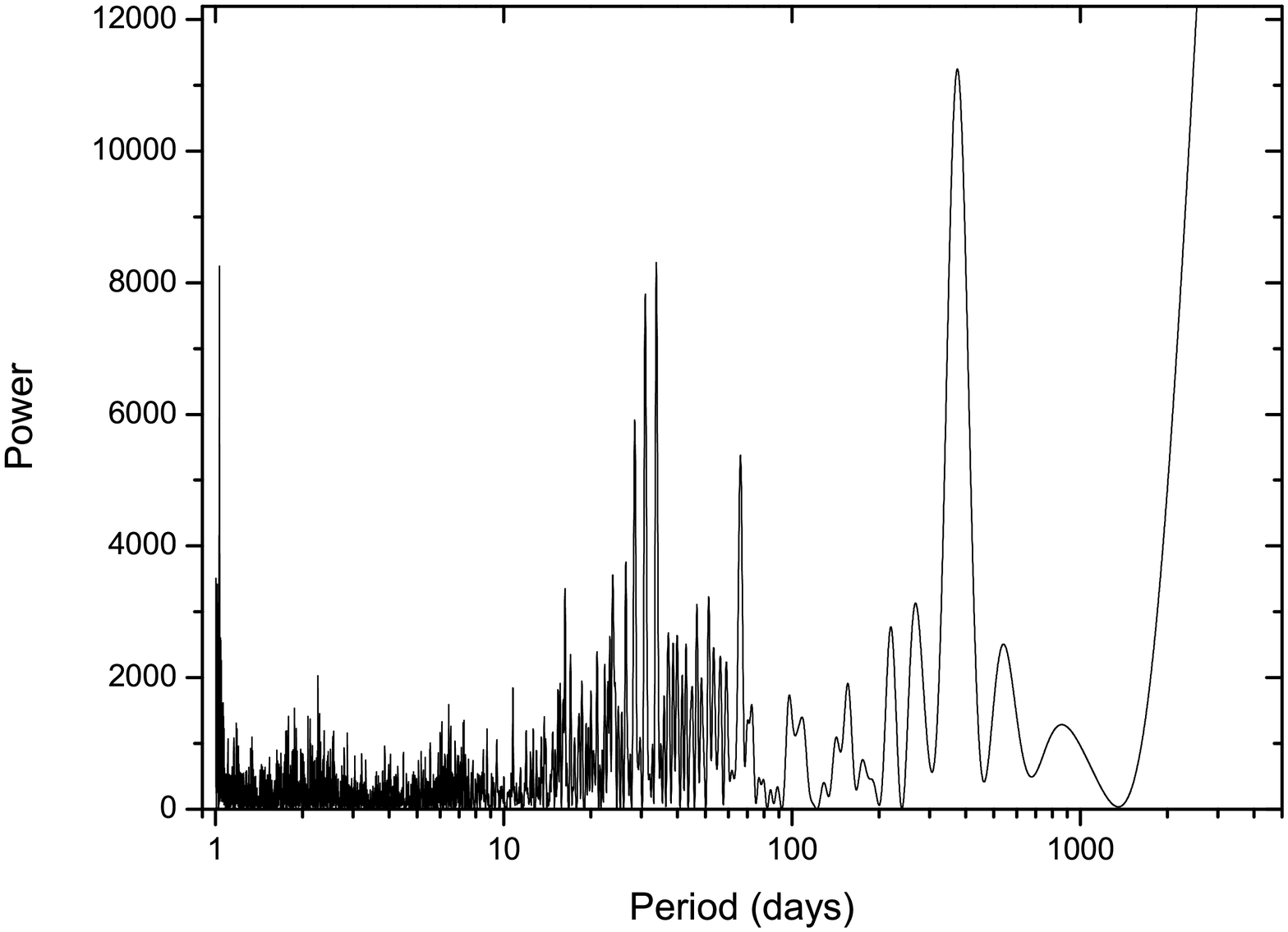}
\caption{Periodograms for HD 181433. Left: LS periodogram of the RV data, a peak at $P = 1171.35$ days is present. The three horizontal lines represent, from top to bottom, the 0.1 \%, 1.0 \% and 10.0 \% FAP levels, respectively. Right: Periodogram of sampling showing peaks that are related with periodicities in the cadence of observations.}
\label{period}
\end{figure*}

We perform a re-analysis of the HARPS data using the Systemic Console\footnote{Available at \url{http://www.oklo.org}} \citep{b16}. Systemic has already been used to derive orbital fits in other works such as \citet{b25} and \citet{b17}. The list of available tools include: Lomb-Scargle (LS) periodogram to identify periodicities in the RV dataset, Lomb-Scargle periodogram of residual to study periodicities in the residual RV dataset, simulated annealing for global multiparameter optimization, while for local multidimensional optimization there are the Levenberg-Marquardt (L-M) scheme which ensures a rapid convergence and the Nelder-Mead (sometimes called AMOEBA) algorithm \citep{b20}.

We held the stellar mass fixed, adopting the value $M_{*} = 0.781 M_\odot $ \citep{b23}. We make the assumption the system is coplanar and viewed edge-on. This conjecture diminishes the quantity of potential orbital configurations greatly, but the plane ($a_{d}$, $e_{d}$) is dynamically representative for the system in the sense that it crosses all resonances \citep{b22}. When long enough time-series of precision data are available, the effects of mutual interactions part of the Newtonian model can potentially help in determining or estimating the masses and inclinations for the planets (see the discussion about this case in Section \ref{newton3}).

\subsection{The Keplerian 3-planet best-fit}

The LS periodogram shows a peak at $P = 1171.35$ days with an estimated false alarm probability (FAP) of $\approx 2 \times 10^{-19}$. Fig. \ref{period} shows the LS periodogram of the full RV data set and the periodogram of sampling. The latest shows peaks that are related with periodicities in the cadence of observations, for instance these can arise from the solar and sidereal day, the synodic month and the solar year.

\begin{figure}
\centering
\includegraphics[width=\columnwidth]{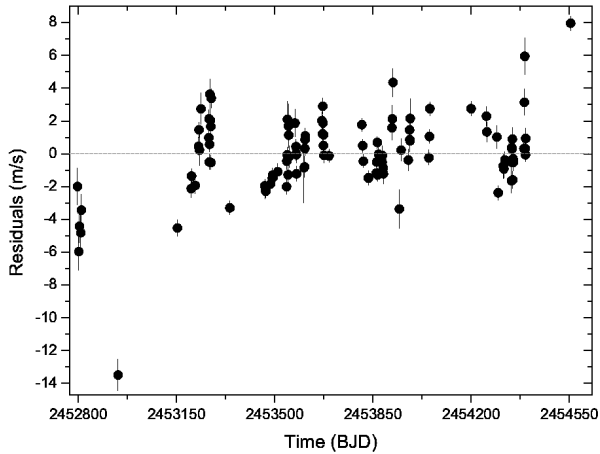}
\caption{Residuals periodogram to the two-planet Keplerian fit.}
\label{resid}
\end{figure}

The residuals periodogram reveals an additional signal at $P = 9.37$ days with FAP of $\approx 3.8 \times 10^{-5}$. The best 2-planet Keplerian fit yields residuals with an rms scatter of 2.44 m/s and reduced chi squared $\chi^2_{red} = 15.7$. The jitter for HD 181433 i.e. the jitter required to have the $\chi^2_{red}$ equal to 1.0, is 2.35 m/s. Fig. \ref{resid} illustrates the periodogram of residuals to the two-planet solution.

To model this long-term trend, we make the starting guess of a planet in an outer 2:1 resonance with planet \textit{c}, we adjust the mass to match the amplitude of the signal and set a small eccentricity. At this point, a Keplerian fit using the L-M algorithm naturally evolves to a solution compatible with the one by \citet{b4}. The best 3-planet fit achieves a $\chi^2_{red} = 4.6$ with an rms scatter of 1.34 m/s and expected jitter of 1.17 m/s. We are not aware of how Bouchy et al. have obtained a lower $\chi^2_{red}$ and a lower value for rms for the same solution.

\begin{figure*}
\includegraphics[width=0.5\textwidth]{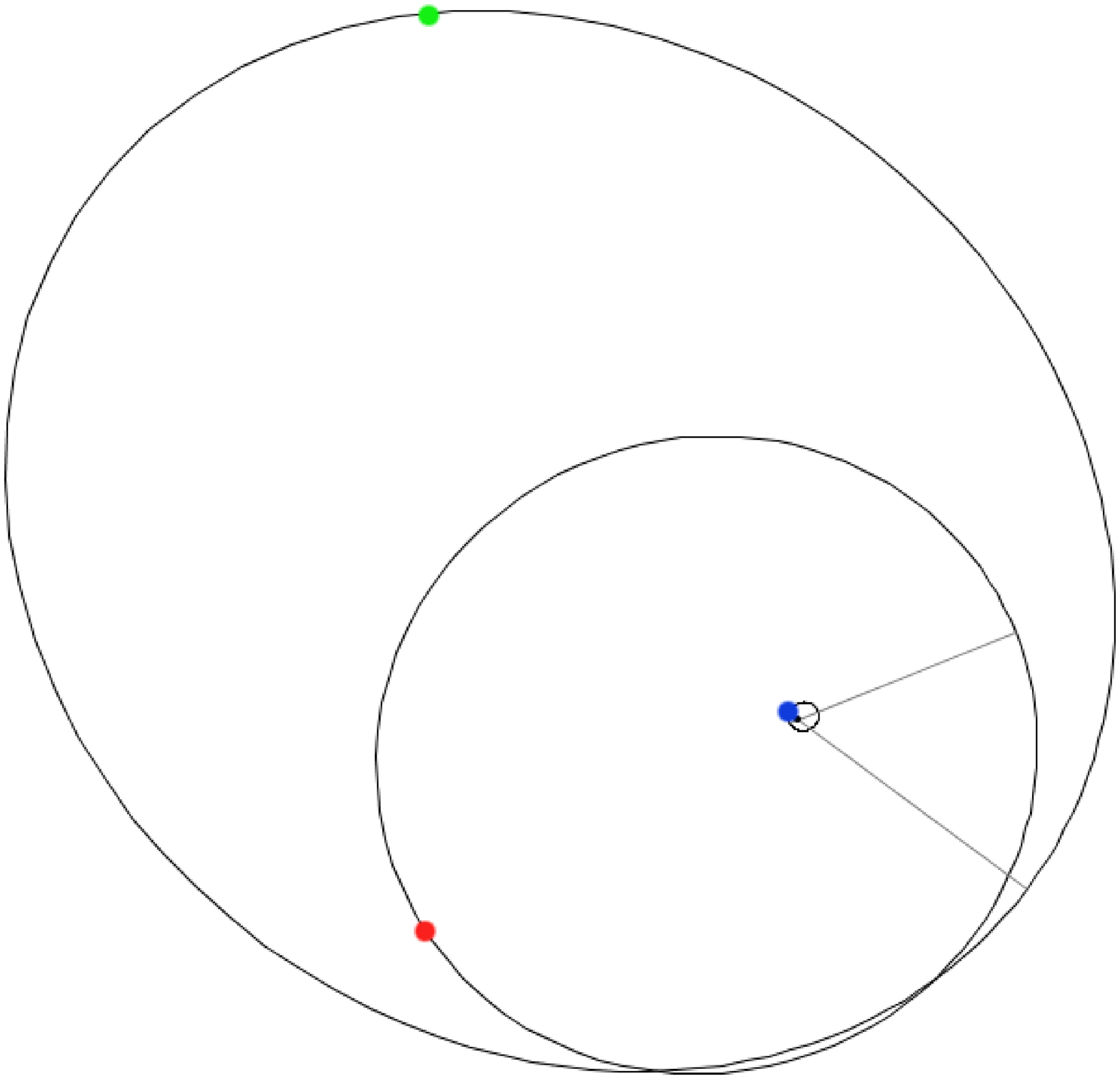}\includegraphics[width=0.5\textwidth]{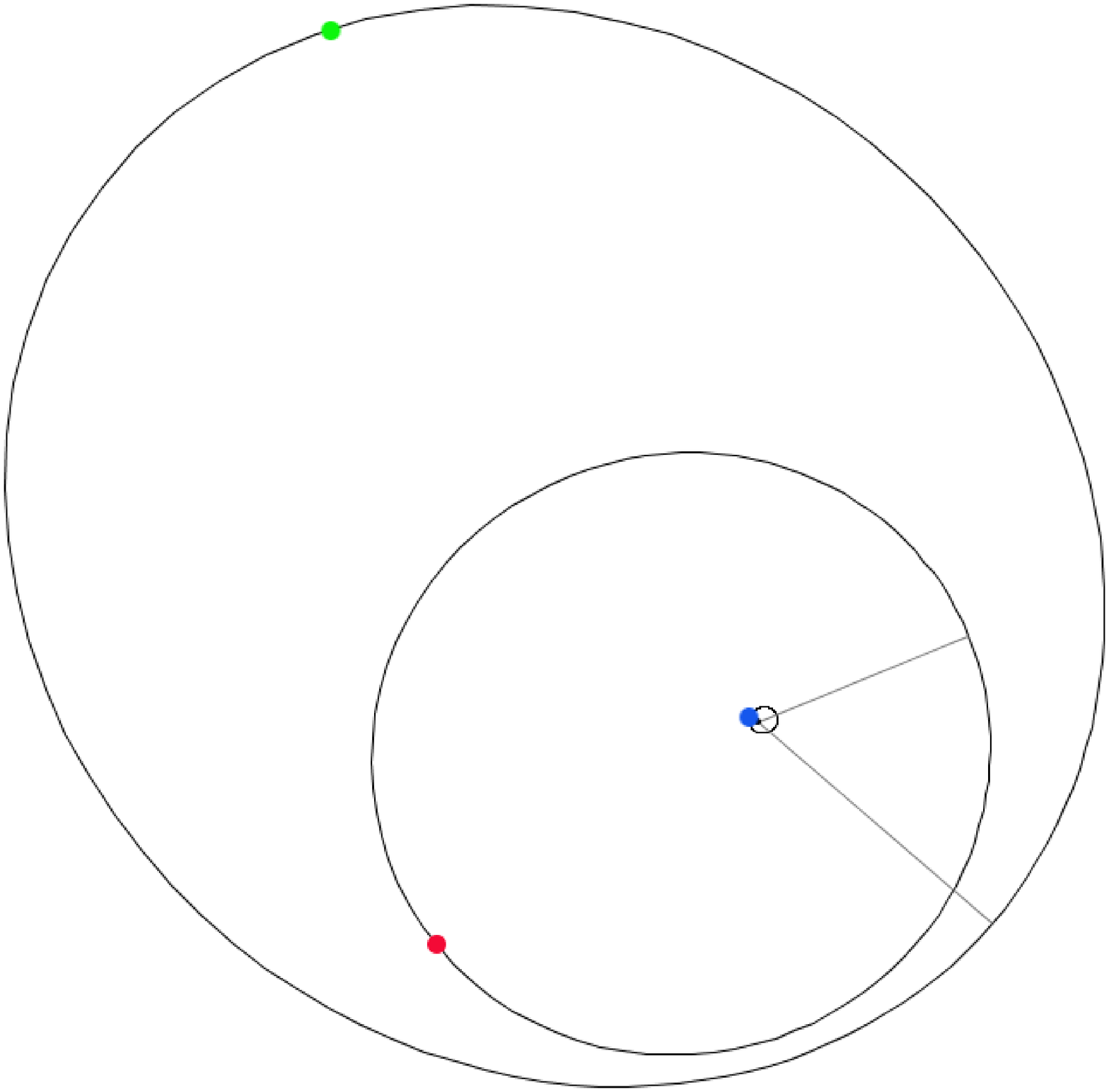}
\caption{Orbital views for the HD 181433 system, the position of the planets along their orbits is the one at the first epoch of observation. The straight lines point toward the pericentres. The osculating elements are valid for the first epoch of observation. Left Panel: Best fit solution, the orbits of planets \textit{c} and \textit{d} cross each other and collisions/ejections occur. Right Panel: Stable best fit solution planets \textit{c} and \textit{d} are in 5:2 MMR.}
\label{orbitalbest}
\end{figure*}

The left panel of fig. \ref{orbitalbest} shows the best-fit orbital configuration at the epoch of the first observation BJD 2452797.87. The orbits of planets \textit{c} and \textit{d} cross each other, because of the strong interactions collisions/ejections occur.

\subsection{The Newtonian 3-planet stable best-fit}
\label{newton3}
Following the argument of \citet{b2} that eccentric orbital solutions can mimic the signal of two planets in 2:1 resonant orbits, we have also tested the hypothesis of a planet in an inner 2:1 resonance with planet \textit{c} but it was not possible to achieve any significant improvement to the goodness of the fit with respect to the 2-planet solution.

The problem of exploring a 16-parameter phase space with stability as additional requirement, can get a first simplification by arguing that the elements of the inner planets are well constrained by observations. In fact, even if we set different starting points for their parameters, the fits for them converge substantially to the same values as these signals are well sampled. A confirmation to this argument comes from the eccentricity of planet \textit{c}, $e_c$, which is a very discriminating parameter toward the stability of the system: if we try to constrain $e_c$ to lower values the fitting we achieve is poor.

Concerning the parameters that describe planet \textit{d}, we notice higher values for the eccentricity are preferred by the fitting. Therefore, at the end the problem can be reduced in finding for each reasonable $P_d$ the largest value for $e_d$ for which planets \textit{c} and \textit{d} do not undergo instability. Likewise, we can say once we have a stable solution for a pair ($P_d$ - $e_d$) we want to investigate if it possible to get a different pair which generates a stable configuration having the same or lower $\chi^2_{red}$. We can describe that as being an empirical Bayesian approach of inferring the stable best-fit rather than a frequentist approach which involves a time consuming number of simulations. Here the investigation is conducted by evidences like the ones given by the collision line (see later on in this Section) and the outcomes of previous simulations.

\begin{figure}
\centering
\includegraphics[width=\columnwidth]{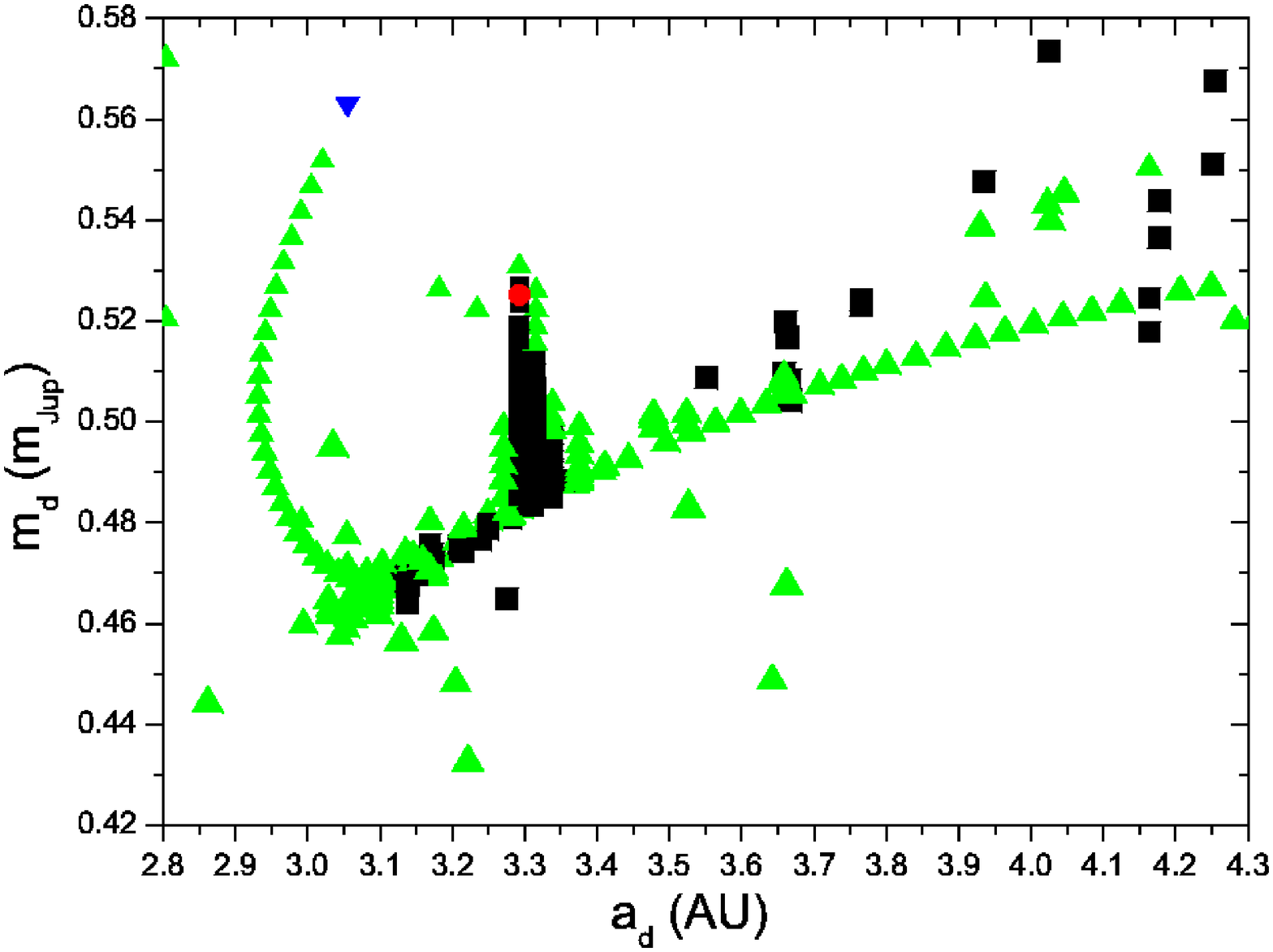} \includegraphics[width=\columnwidth]{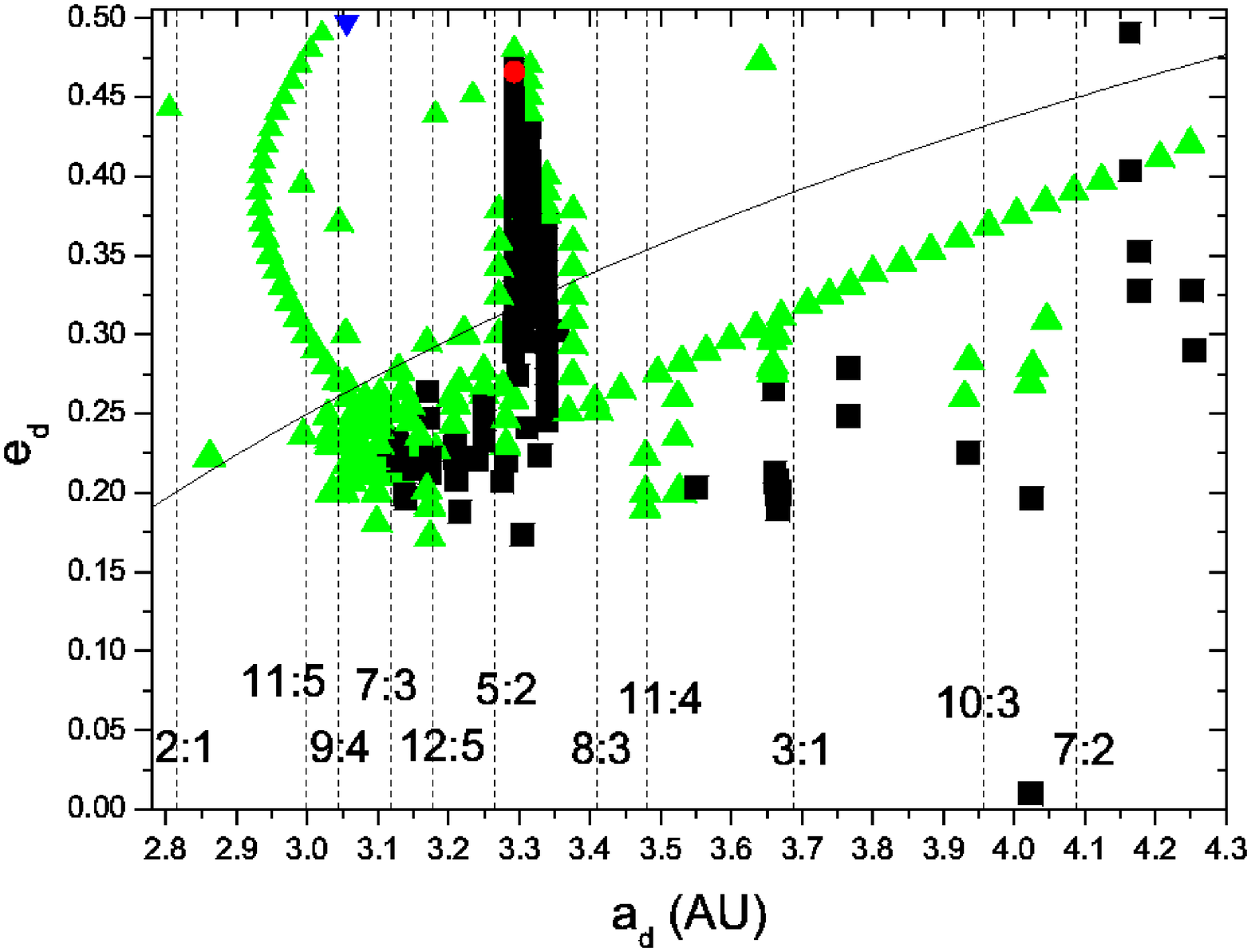}
\caption{The best fits obtained for the HD 181433 planetary system. Top panel: In terms of the mass and semi-major axis of the outermost planet. Bottom panel: In terms of the eccentricity and semi-major axis of planet \textit{d}; the collision line is depicted, the nominal positions of the most relevant MMR are also labelled and marked by dashed lines. The statistically best-fit is indicated with an inverted triangle (blue), the stable best-fit is denoted with a circle (red). With triangles (green) we represent unstable configurations while with squares (black) we refer to models stable for at least 1 Myrs. The size of each symbol is proportional to its $\chi^2_{red}$, i.e. smaller symbols indicate better fits. The 5:2 and 7:2 MMRs retain stability even for values of $e_d$ over the collision line.}
\label{chivsmd}
\end{figure}

The second and third planet reside in regions spanned by a number of strong low-order MMRs (see Fig. \ref{chivsmd} later on). We are aware of the protective role of some MMR. For instance, the 2:1 MMR associated with the SAR consent together stable configurations even for enormously high eccentricities, $\simeq$ 0.95-0.98 (Go\'zdziewski et al. 2003 and references therein). This could explain a very large eccentricity for planet \textit{d} and still preserve the system stability by keeping the planets away from close encounters. Actually, a modification of the relative phase of the planets strongly affects the synthetic RV curve and a \textit{stable} resonance configuration can be far from being consistent with the RV observations. We find that manipulating the values of $\omega_d$ and of the mean anomaly, $M_d$, to get stable configurations is highly unfavourable by the RV data (i.e. poor fits are obtained). Therefore, this supports the argument arisen in the previous paragraph about performing an exploration focused on the ($P_d$ - $e_d$) space while leaving to the algorithm the task of fitting, without constraints of any sort, the other parameters and in particular $m_d$, $\omega_d$ and $M_d$.

To perform Newtonian orbital fits, Systemic offers different method such as the Runge-Kutta, Hermite $4^{th}$ order and Gragg-Bulirsch-Stoer integrators. Fitting a Newtonian solution takes longer than a Keplerian model but it assures short time scale interactions, relevant for planets \textit{c} and \textit{d}, are considered.

The following step is studying the stability of each distinct fit over a period of time related to the time-scale of unstable behaviours. For these long-term evolution tests, direct N-body integrations are applied to the orbital solutions considered. We integrate the orbits for at least 1 Myrs using the Wisdom-Holman Mapping integrator available in the SWIFT software package \citep{b10}. We use a time step approximately equal to a twentieth ($\approx$ 1/20) of the orbital period of the innermost planet. To study close encounters, we use the available Bulirsch-Stoer integrator with a tolerance parameter of $10^{-9}$. We identify each configuration as being a stable system if orbits stay well bounded over an arbitrarily long period of time.

\begin{figure}
\centering
\includegraphics[width=\columnwidth]{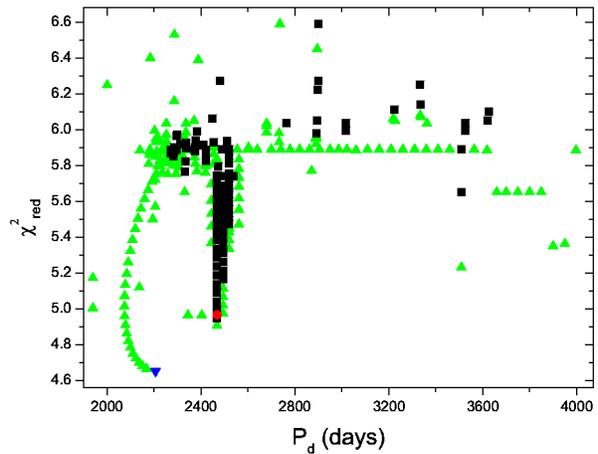}
\caption{The outcome of the simulations for the HD 181433 planetary system in terms of the statistical goodness of the fits and the orbital period of the outermost planet. The statistically best-fit is indicated with an inverted triangle (blue), the stable best-fit is denoted with a circle (red). With triangles (green) we represent unstable configurations while with squares (black) we refer to models stable for at least 1 Myrs. Stable configurations do not exist in the near neighbourhood of the statistically best-fit,  the deep minimum with stable models represents the region where a 5:2 MMR configuration is possible.}
\label{chivspd}
\end{figure}

The results of our analysis are outlined in Figs. \ref{chivspd} and \ref{chivsmd}. Here we label as stable the solutions that survived at least for 1 Myrs. Once again, we fix ($P_d$ - $e_d$) and then look for the best fit, starting the L-M scheme with initial points derived from previously studied configurations. The L-M algorithm and Amoeba offer a clear representation of the parameter space. The dynamical analysis reveals a narrow and long band around 3.3 AU and a small island around 3.2 AU where good fitness is achieved and stability requirements are met. We find a configuration, that we label as stable best-fit, which survives for at least 250 Myrs (see Sect. \ref{longterm} for an in-depth examination). Other models scored a better $\chi^2_{red}$ but did not preserve stability for the same amount of time. Therefore, the stable best-fit seems to lie on the border of a chaotic and unstable zone where small changes on the parameters of the outermost planet may push the system into a strongly chaotic state leading, in some scenarios, even to its disruption.

Figure \ref{chivspd} illustrates how stable configurations do not exist in the near neighbourhood of the statistically best-fit; smaller quantities for $e_d$ are needed in order for the models to retain stability and that increases the value of $\chi^2_{red}$.

The top panel of Figure \ref{chivsmd} shows the best fits obtained during our investigation in terms of the mass for planet \textit{d}, $m_d$, and the semi-major axis $a_d$. The picture makes clear how to explain a certain RV amplitude $K_d$, a bigger mass $m_d$ is required as long as $a_d$ increases.

The bottom panel of Figure \ref{chivsmd} illustrates the results of our analysis in the semi-major axis-eccentricity plane ($a_d$ - $e_d$). The parameters represented are the osculating elements at the epoch of the first observation. We show the collision line which is defined in terms of semi-major axes and eccentricities as $a_c(1+e_c)=a_d(1+e_d)$. This line denotes the region where the mutual interactions of relatively massive companions can rapidly destabilize the configuration and is calculated for $e_c = 0.269$ and $a_c = 1.773$ AU (the values are from our stable best-fit solution, see Table \ref{param} later on). Note how the statistically best-fit is positioned well over the collision line. We also identify the most relevant MMRs between planets \textit{c} and \textit{d}, such as the 2:1, 11:5, 9:4, 7:3, 12:5, 5:2, 8:3, 11:4, 3:1, 10:3 and 7:2. The position of the indicated locations has been calculated with respect to the values of the stable best-fit. Planets in some resonant configurations, even if under the collision line, exchange angular momentum rapidly; their eccentricities are quickly pumped and that may lead again to instabilities and self-disruptions. In particular, we have found models that show a stable and bounded evolution for many Myrs before the unstable behaviours are manifested. On the contrary, other resonant configurations, such as the 5:2 and 7:2, are observed to retain stability even for values over the collision line.

\begin{figure*}
\includegraphics[width=0.5\textwidth]{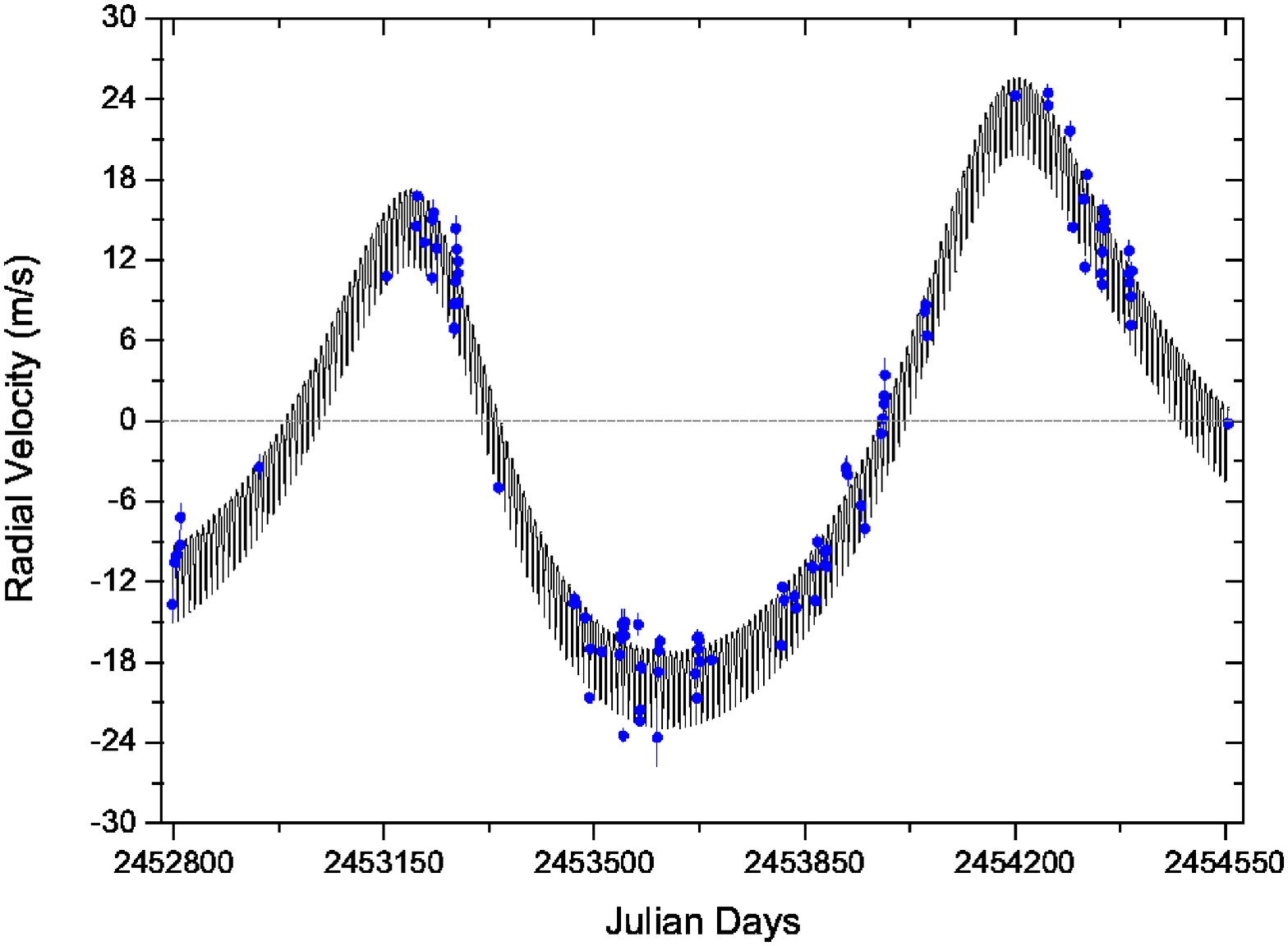} \includegraphics[width=\columnwidth]{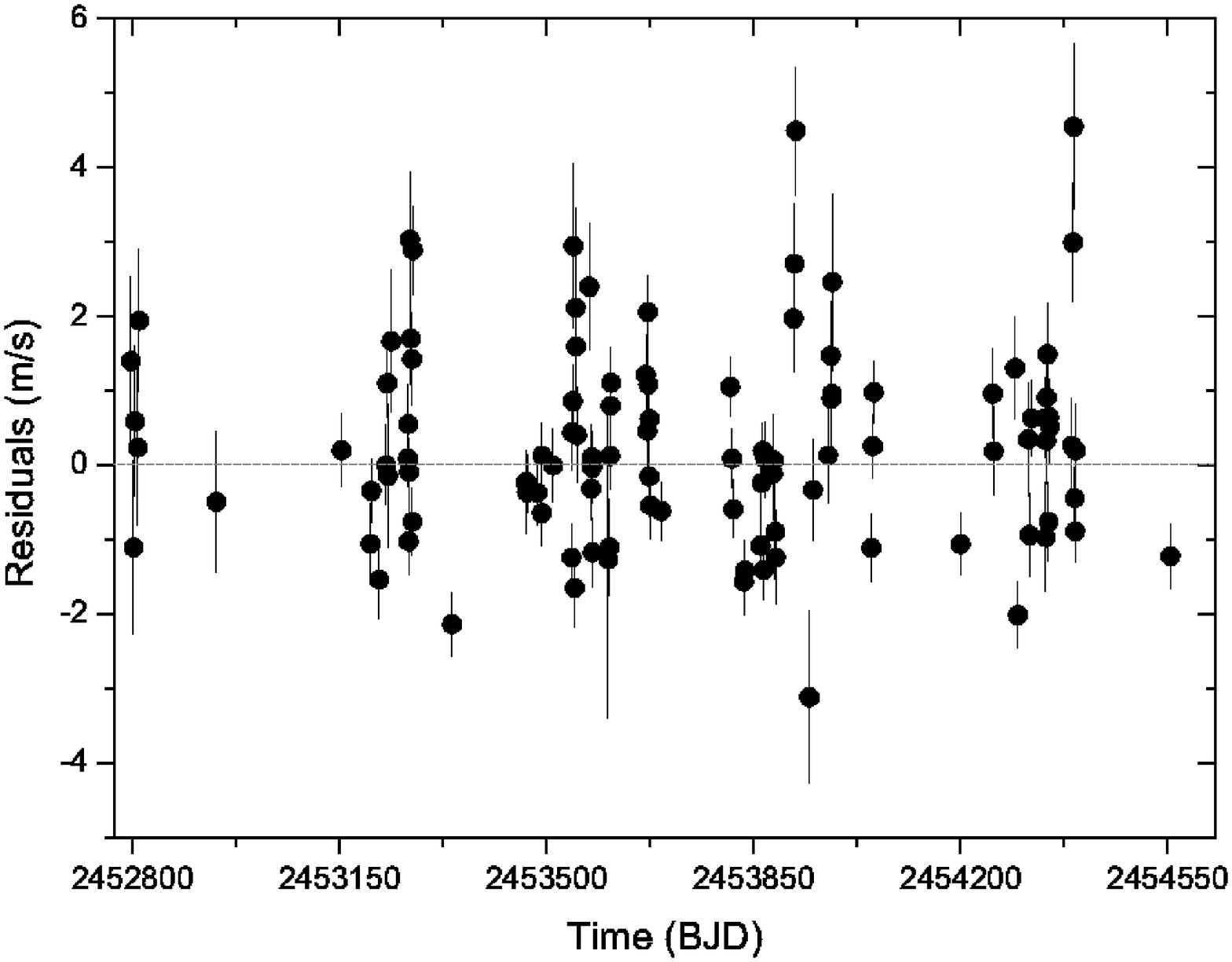}
\caption{The Newtonian 3-planet stable best-fit model and residuals periodogram for the HD 181433 RV data.}
\label{resid3}
\end{figure*}

Table \ref{param} reports the determined set of orbital elements for the stable best-fit. For each planet, we list period (\textit{P}), time of periastron passage ($T_{peri}$), eccentricity (\textit{e}), argument of pericentre ($\omega$), semi-amplitude (\textit{K}), minimum mass ($m\sin i$) and semi-major axis (\textit{a}); we indicate also the stellar offset (\textit{V}). This model has $\chi^2_{red} = 4.96$, an rms scatter of 1.36 m/s and the expected jitter of HD 181433 is 1.19 m/s. Figure \ref{resid3} displays the RV data fitted to this model along with the residuals. The right panel of figure \ref{orbitalbest} shows the orbital configuration of the system, this time the orbits of planets \textit{c} and \textit{d} do not cross each other.

Since the stable best-fit is found in an active region, rather than estimating an uncertainty on each parameter, we think Figs. \ref{chivspd} and \ref{chivsmd} are more useful in visualizing the results of the dynamical study and highlight what is plausible to expect from new observations. If we compare our results with what has already been published for this planetary system\footnote{\citet{b4} do not report directly the uncertainties for the masses and semi-major axes, in this case we considerer what is available on \url{http://exoplanets.org}}, we find that the parameters of planet \textit{b} and \textit{c} are confirmed to be already well constrained with just $K_c$ not compatible within the 3 $\sigma$. For planet \textit{d}, all the elements are found within the 3 $\sigma$ from the original conclusion. However, it is worth to underline how to explain the very large eccentricity of the third planet and to retain a good fit to the present data, the uncertainity on the location of planet \textit{d} reduces drammatically to the narrow band where the 5:2 MMR is possible. Hence, this supports how a dynamical study can be fundamental in interpreting observations, producing a self-consistent model compatible with the data and giving substantial constraints on the orbital parameters. 

\begin{table*}
 \centering
 \begin{minipage}{140mm}
  \caption{Orbital and physical parameters of the stable best-fit found for the HD 181433 planetary system. The osculating elements are given for the epoch of the first observation BJD 2452797.8654.}
\centering
  \begin{tabular}{@{}lccc@{}}
  \hline
   \textbf{Parameter} & \textbf{HD 181433 b} & \textbf{HD 181433 c} & \textbf{HD 181433 d} \\
 \hline
 P (days) & 9.37459 & 975.41 & 2468.46 \\
$T_{peri}$ (BJD-2450000) & 2788.9185 & 2255.6235  & 1844.4714 \\
e & 0.38840 & 0.26912 & 0.46626 \\
$\omega$ ($^{\circ}$) & 202.039 & 22.221 & 319.129 \\
K (m/s) & 2.57 & 14.63 & 9.41 \\
$m\sin i$ ($m_{JUP}$) & 0.02335 & 0.65282 & 0.52514 \\
$m\sin i$ ($m_\oplus$) & 7.4 & 207.5 & 166.9 \\
a (AU) & 0.08013 & 1.77310 & 3.29347 \\
\hline
V (m/s) & & 40212.846 & \\
rms (m/s) & & 1.36 & \\
$\chi^2_{red}$ & & 4.96 & \\
\hline
\label{param}
\end{tabular}
\end{minipage}
\end{table*}

The data do not offer any possibility of constraining the orbital inclinations. The Newtonian model cannot be particularly improved because, aside from the fact the signal of the outer planet is not well sampled, we need to wait for secular timescales before the variations in \textit{i} can be spotted via the RV method. In fact, we notice that for planets GJ 876 \textit{b} and \textit{c} which have the strongest mutual gravitational interactions, more than 11 years of observations (corresponding to more than 60 orbits of the outer planet) were used to give a reasonable estimate of the inclinations \citep{b5}.

\subsection{Additional planets?}
\label{4pl}
\begin{figure}
\centering
\includegraphics[width=\columnwidth]{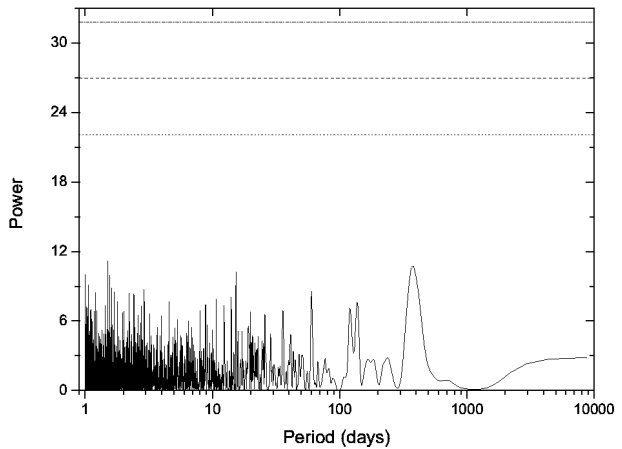}
\caption{Periodogram of the residuals to the 3-planet solution for HD 181433. It does not display any strong peak that would support the evidence for additional planets in the system. The three horizontal lines represent, from top to bottom, the 0.1 \%, 1.0 \% and 10.0 \% FAP levels, respectively.}
\label{resi3}
\end{figure}

Finally, following the argument that the proximity of the best fit to the collision line may indicate the presence of further planets \citep{b9}, we aim to search for 4-planet Newtonian solutions. The periodogram of the residuals to the 3-planet solution, in Fig. \ref{resi3}, displays no strong peaks that would support the evidence for additional planets in the system. Apart from more distant companions, in the inner region of the system a terrestrial planet can only survive if located between planets \textit{b} and \textit{c}. In fact, already planet \textit{b} is found in the proximity of the parent star and the area between planets \textit{c} and \textit{d} is dominated by the strong interactions that interest the two giant planets in eccentric orbits. The existance of this last planet would support the ``packed planetary systems'' hypothesis \citep{b3}.

The present data do not allow making any supposition about possible outer planets (for example an object at 7 AU would have an orbital period of around 7700 days). On the other hand, with a super-Earth in the stable zone between planets \textit{b} and \textit{c} the fit improves. However, this signal would be at the noise level with an F-test of the order of 30\%. The F-test indicates the probability that a planetary model would produce a signal similar to the one due just to noise fluctuations in the data (Marcy et al. 2005 and references therein), so additional observations are required to investigate on the presence of a super-Earth or less massive planet in this stable region.

\section[]{LONG-TERM BEHAVIOUR OF THE STABLE BEST-FITTING CONFIGURATIONS}
\label{longterm}
Because of the proximity of the two outermost planets, the system cannot be stable unless a resonant mechanism is present to avoid close encounters. In this Section we aim to deepen the study of the stable best-fit configuration as well as investigating the evolution of the orbital elements, the secular resonant arguments and critical angles of some particular configurations consistent with the RV observations.

\begin{figure*}
\includegraphics[width=0.5\textwidth]{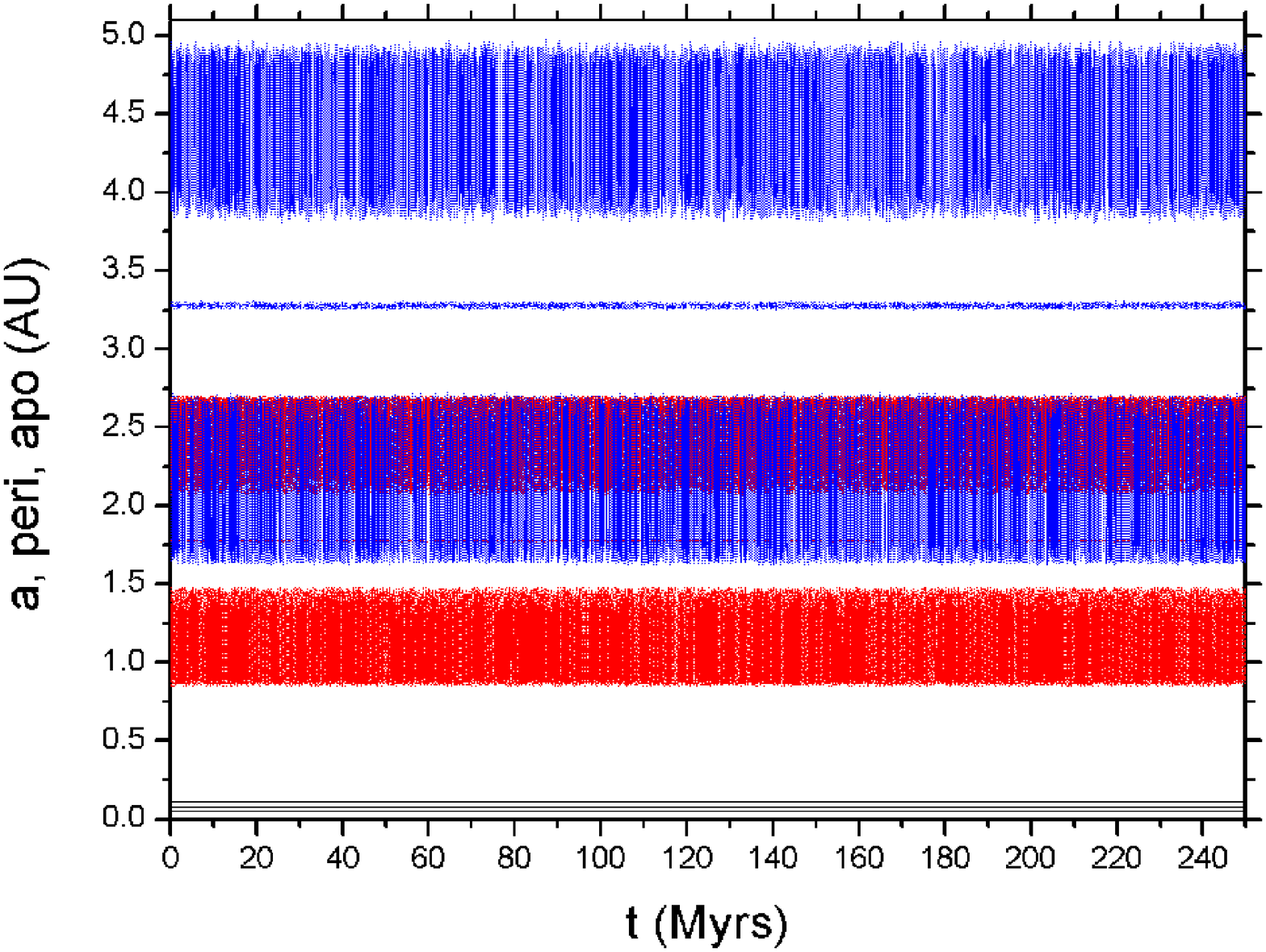}\includegraphics[width=0.5\textwidth]{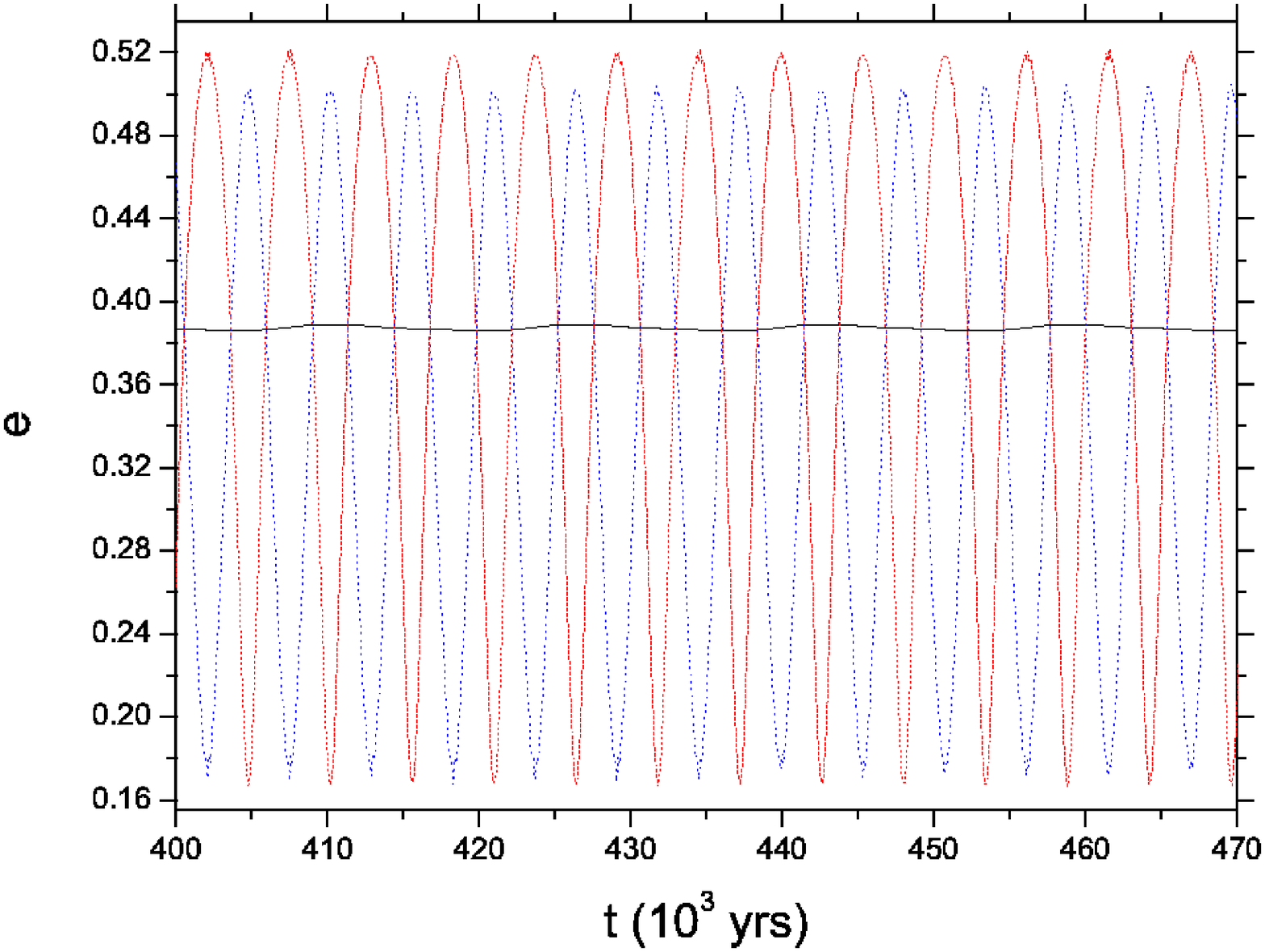}
\includegraphics[width=0.5\textwidth]{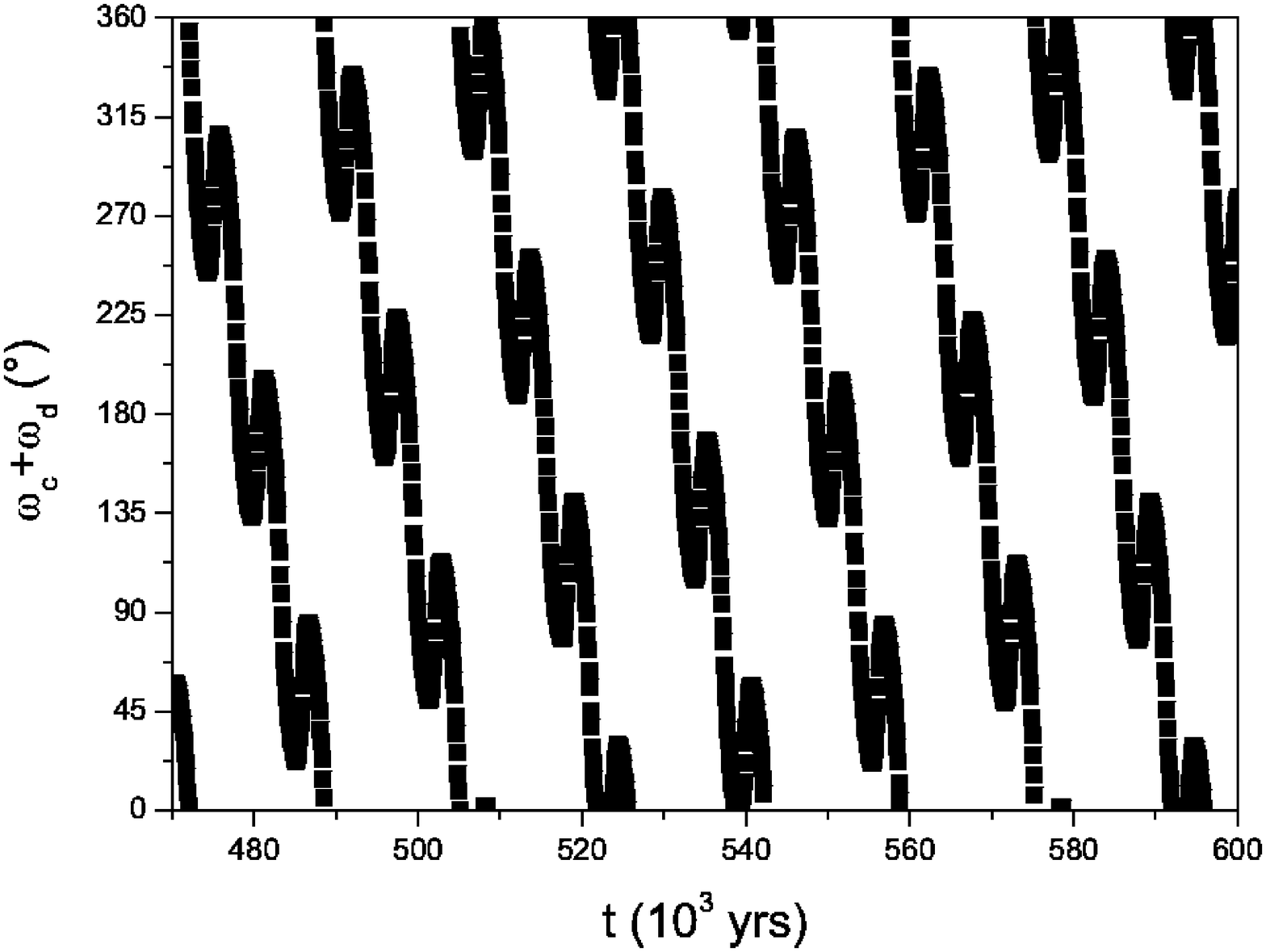}\includegraphics[width=0.5\textwidth]{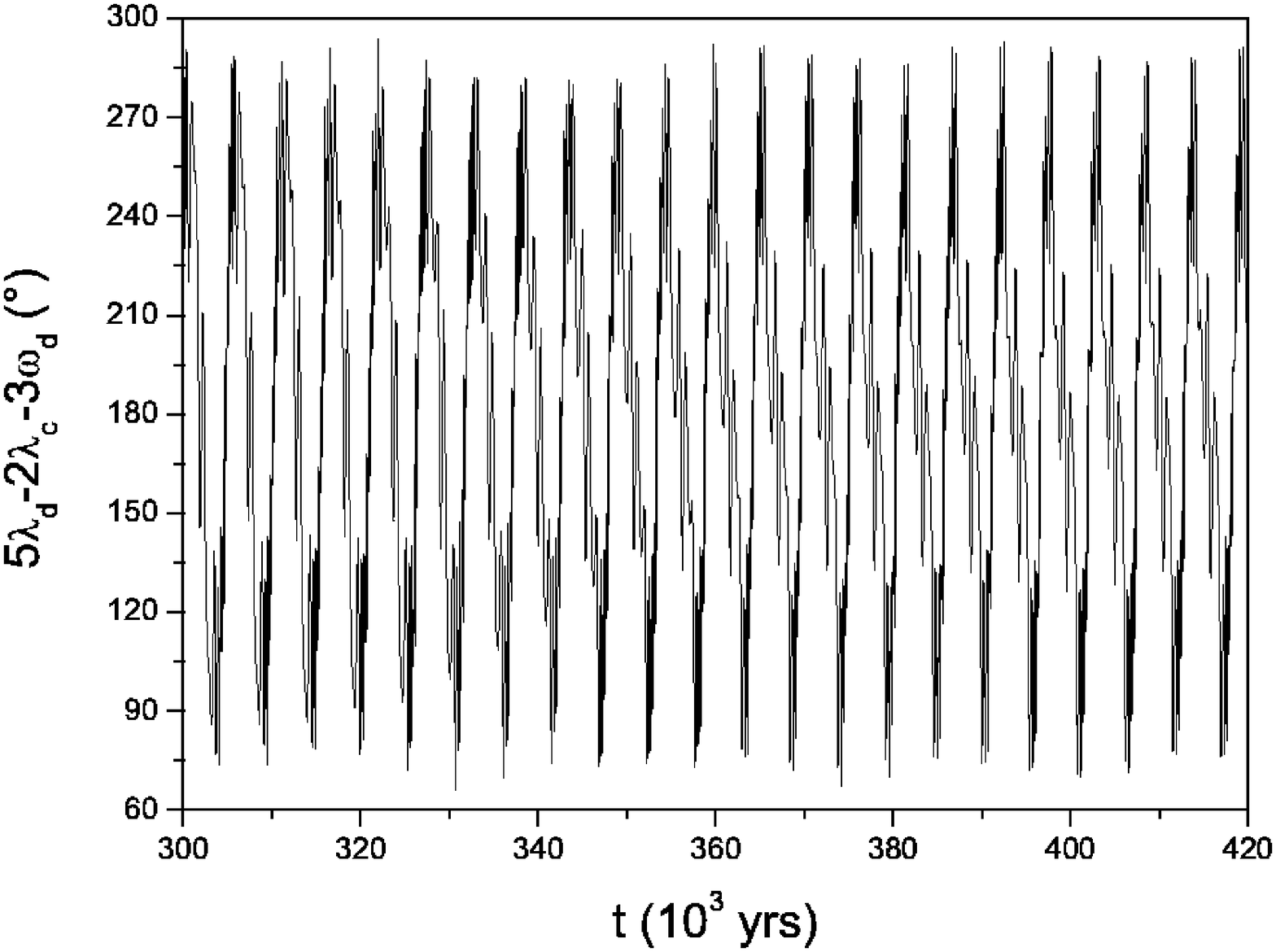}
\caption{Evolution of some orbital elements for the stable best-fit configuration. Top panels: The time evolution of the semi-major axes and of the eccentricities are depicted, the solid line (black) indicates planet \textit{b}, the dashed line (red) denotes planet \textit{c} and the dotted line (blue) represents planet \textit{d}. $e_c$ moves in the interval 0.17-0.52 while $e_d$ in the range 0.17-0.50. Bottom panels: A secular resonant angle and a critical argument of the 5:2 MMR are illustrate.}
\label{orbelbest}
\end{figure*}

For the stable best-fit, Fig. \ref{orbelbest} shows in the subsequent panels the time evolution of the semi-major axes and of the eccentricities. Moreover, a secular resonant angle and a critical argument of the 5:2 MMR are also illustrated. Specifically, the top-left panel in Figure \ref{orbelbest} highlights that for the 250 Myrs of the numerical integration the apocentre of planet \textit{c} and the pericentre of planet \textit{d} share the same region. If we get a close-in view of the situation, we notice that actually they never cross each other. In particular, the pericentre of \textit{d} is internal to the apocentre of \textit{c}. The former approaches the value of $a_d$ around every 50000 years. It is probably a resonance that, protecting the companions from close encounters, allows the stability of the system. The top-right panel illustrates the relative large range in which $e_c$ and $e_d$ evolve. The peak-to-peak amplitude is covered in around 2500 years only. $e_c$ moves in the interval 0.17-0.52 while $e_d$ in the range 0.17-0.50. The present eccentricities fall in the middle of these intervals indicating that the system has been snapped in a statistically quite probable state. Also, such a large range reminds that in multiple-planet systems the orbital eccentricities can vary considerably through secular interactions on timescales that are long compared to observational baselines but short compared to the age of the systems. Therefore, when doing statistical studies on exoplanetary systems the planetary orbits should normally be described by a complete distribution of values for the eccentricities rather than just by the present quantities (see also Adams \& Laughlin 2006). This model is not observed to be in SAR. The bottom-left panel indicates the time evolution of the secular argument $\omega_c+\omega_d$ which alternates librations with circulations. Besides, we find that $5n_d-2n_c \approx -3.4 ^{\circ}/yr$ indicating the proximity of the 5c:2d MMR, therefore here we have a scenario similar to the Jupiter-Saturn case in the Solar System. We have studied the resonant arguments of this MMR and found that the angle $5\lambda_d-2\lambda_c-3\omega_d$ librates around 180$^{\circ}$ with a semiaplitude of about 110$^{\circ}$. Thus, this configuration is seen to be locked in a MMR, the bottom-right panel illustrates how this critical angle evolves with time.

The mass of planet \textit{b} is negligible with respect \textit{c} and \textit{d} so we can assume the dynamics of the two
giants is not disturbed much by the presence of the rocky planet close by. Then, we study some
possible configurations with planets \textit{c} and \textit{d} near MMR. We do not find any plausible ($\chi^2_{red} \leq 6.03$)
stable solution that would correspond to the 2:1, 11:5, 9:4, 7:3, 8:3, 11:4 and 10:3 MMRs. In
particular, the configurations nMMR 11:5, 9:4 and 7:3 seem preferred by the data but the excessive
pumping of the eccentricities causes close encounters and planetary scatterings which do not favour
stability. On the other hand, outer MMRs are possible because here in particular the planets are
more spread and collisions can be avoided.

\begin{figure*}
\includegraphics[width=0.5\textwidth]{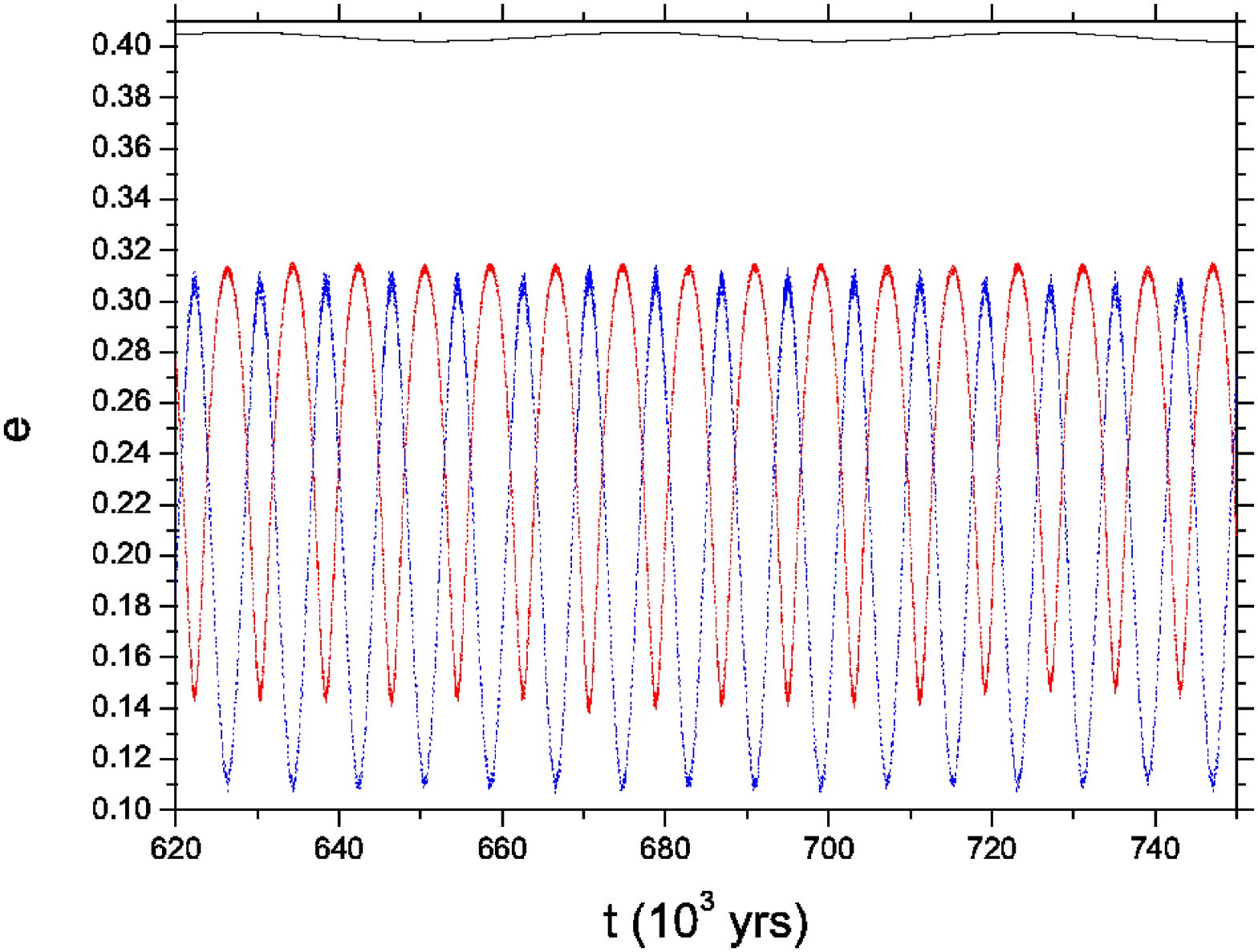}\includegraphics[width=0.5\textwidth]{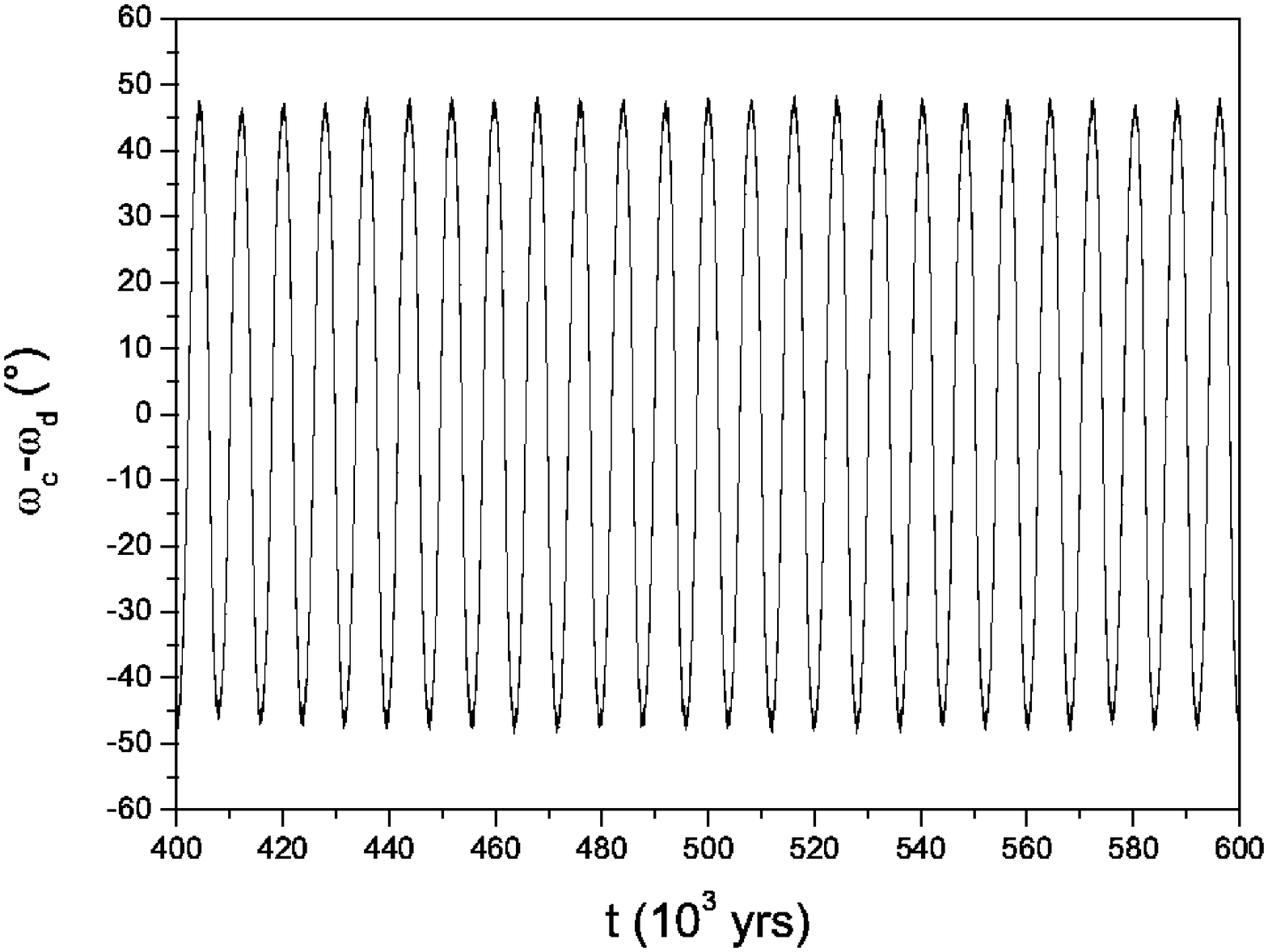}
\includegraphics[width=0.5\textwidth]{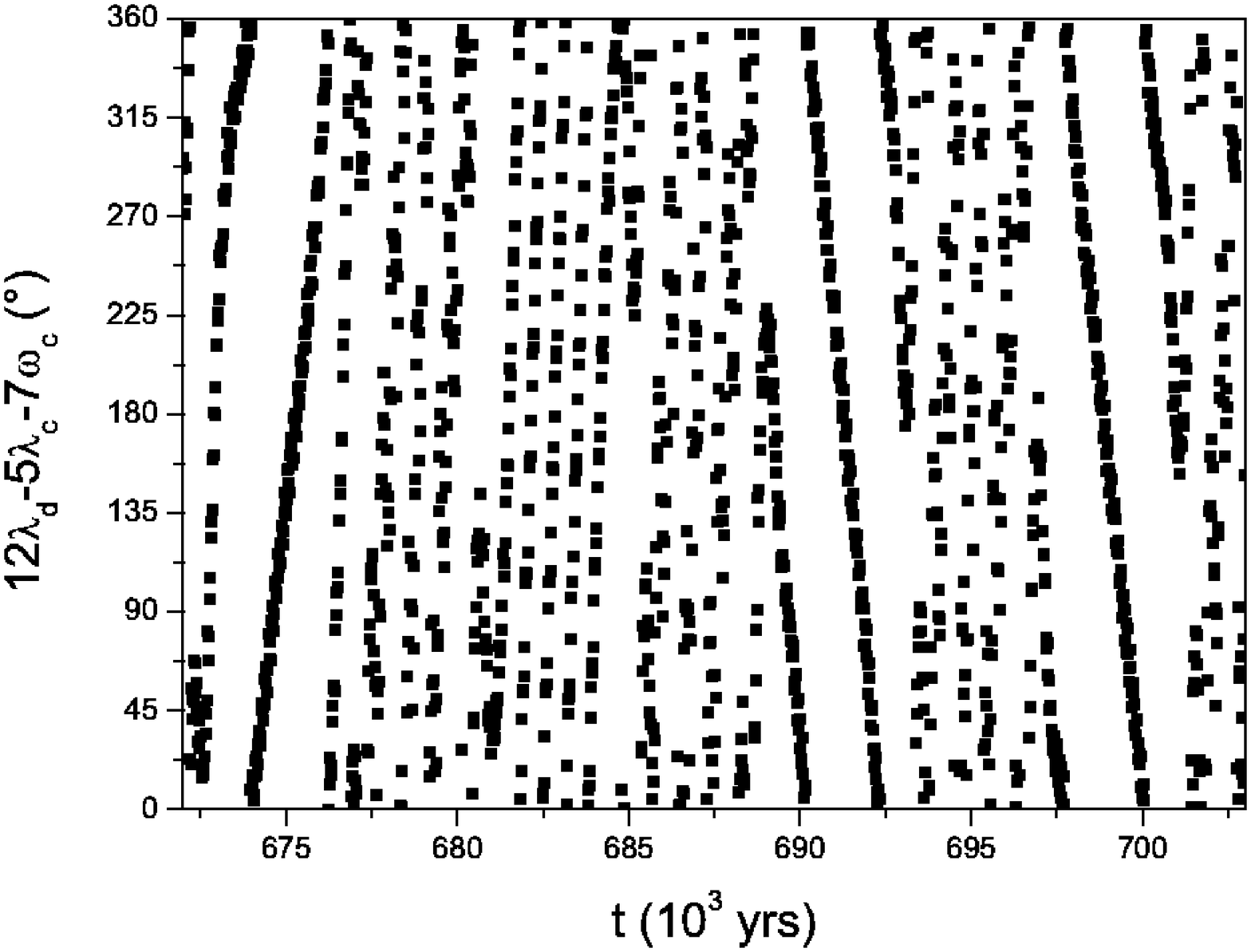}\includegraphics[width=0.5\textwidth]{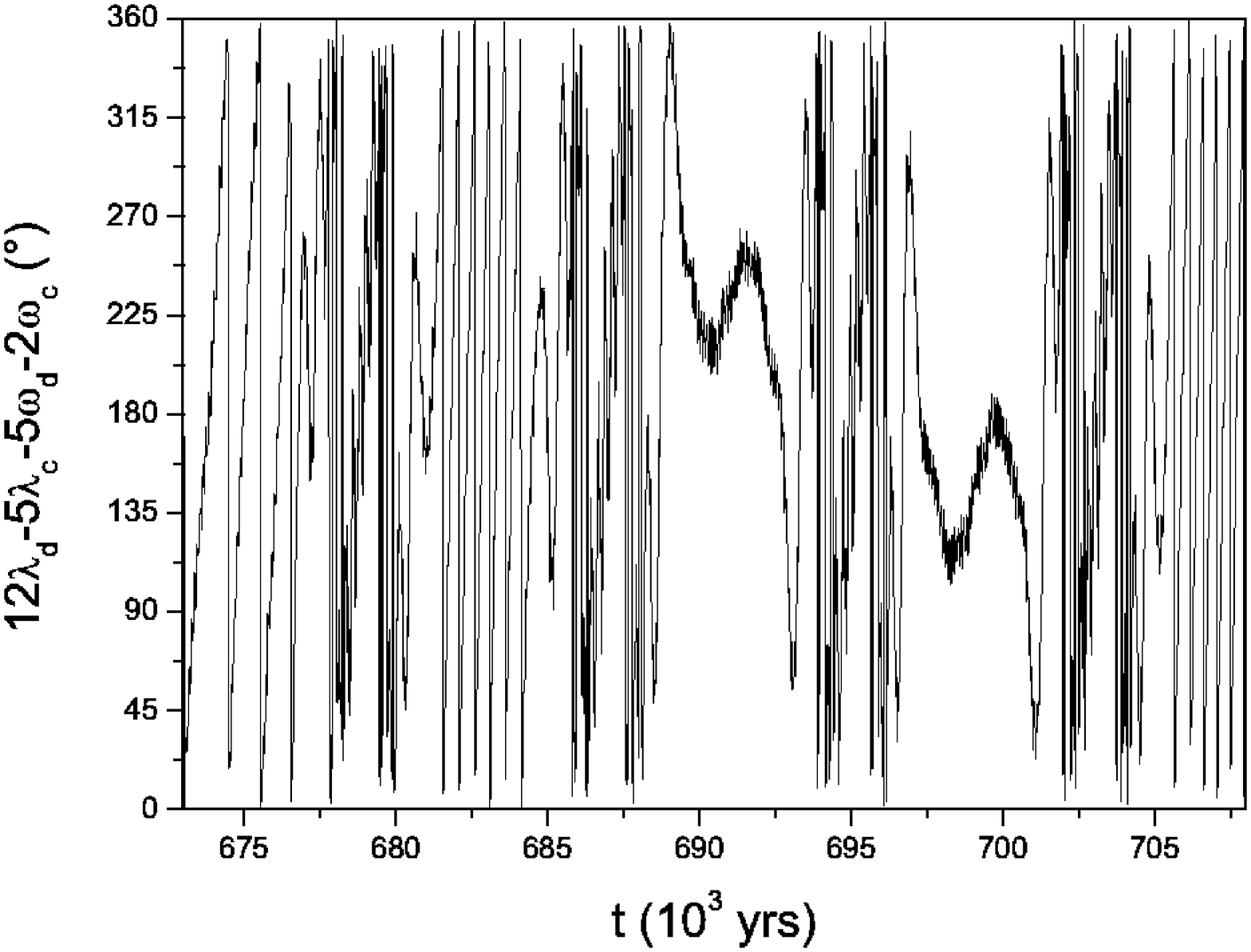}
\caption{Evolution of some orbital elements for the nMMR 12:5 configuration. Top-left: The solid line (black) indicates planet \textit{b}, the dashed line (red) denotes planet \textit{c} and the dotted line (blue) represents planet \textit{d}. $e_c$ moves in the interval 0.14-0.31 while $e_d$ in the range 0.11-0.31. This model is in SAR.}
\label{125}
\end{figure*}

\begin{figure*}
\includegraphics[width=0.33\textwidth]{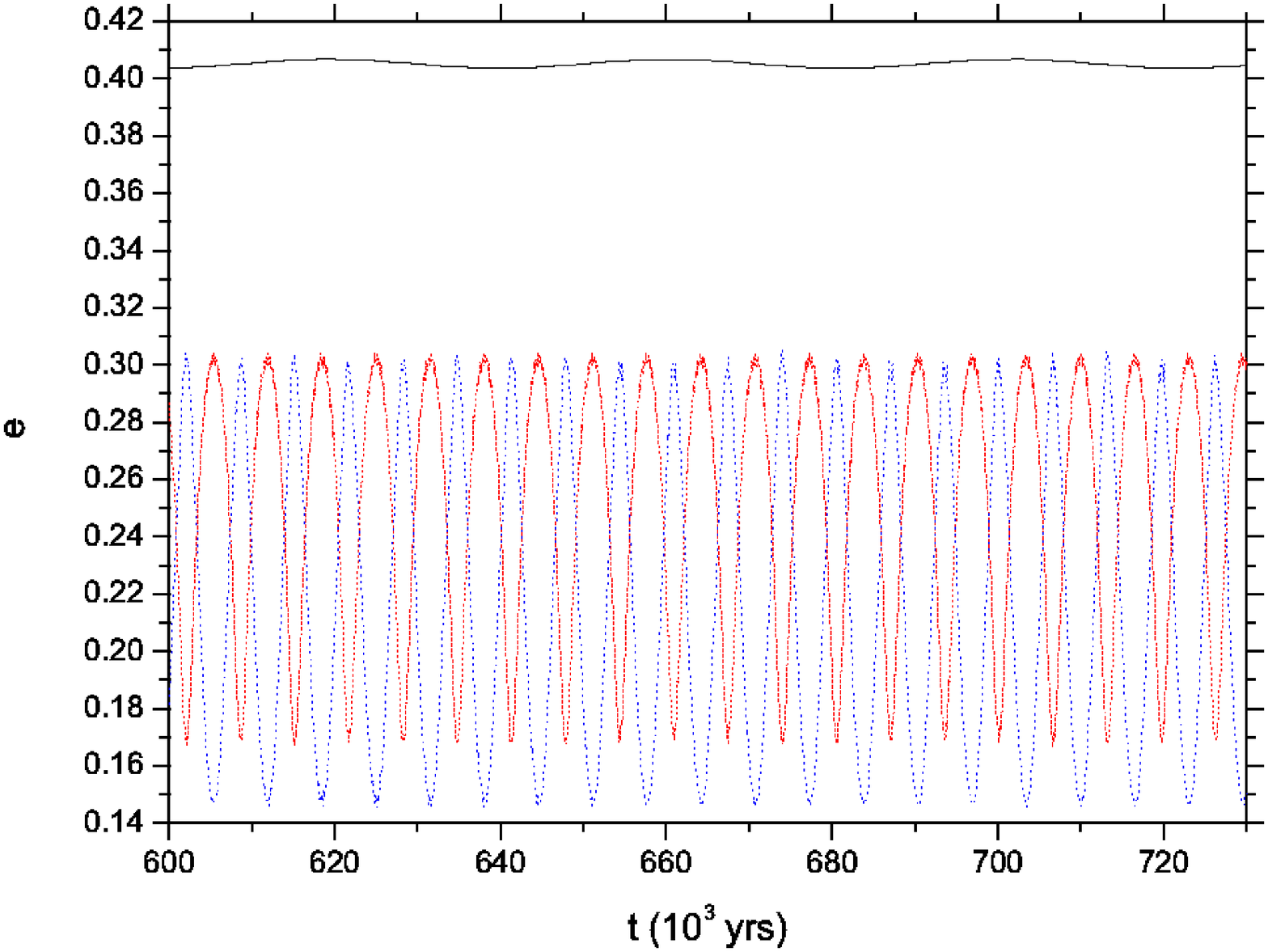}\includegraphics[width=0.33\textwidth]{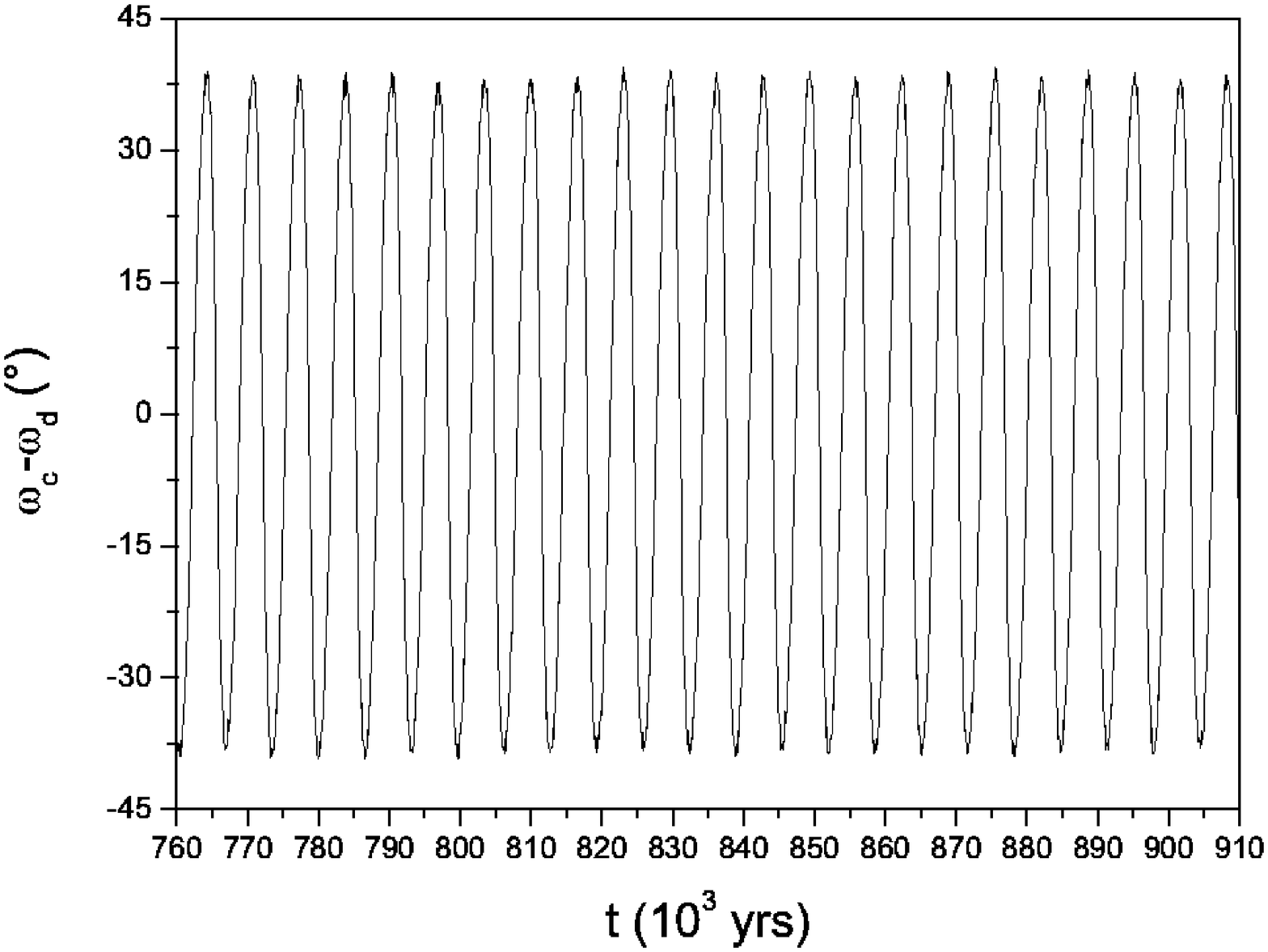}\includegraphics[width=0.33\textwidth]{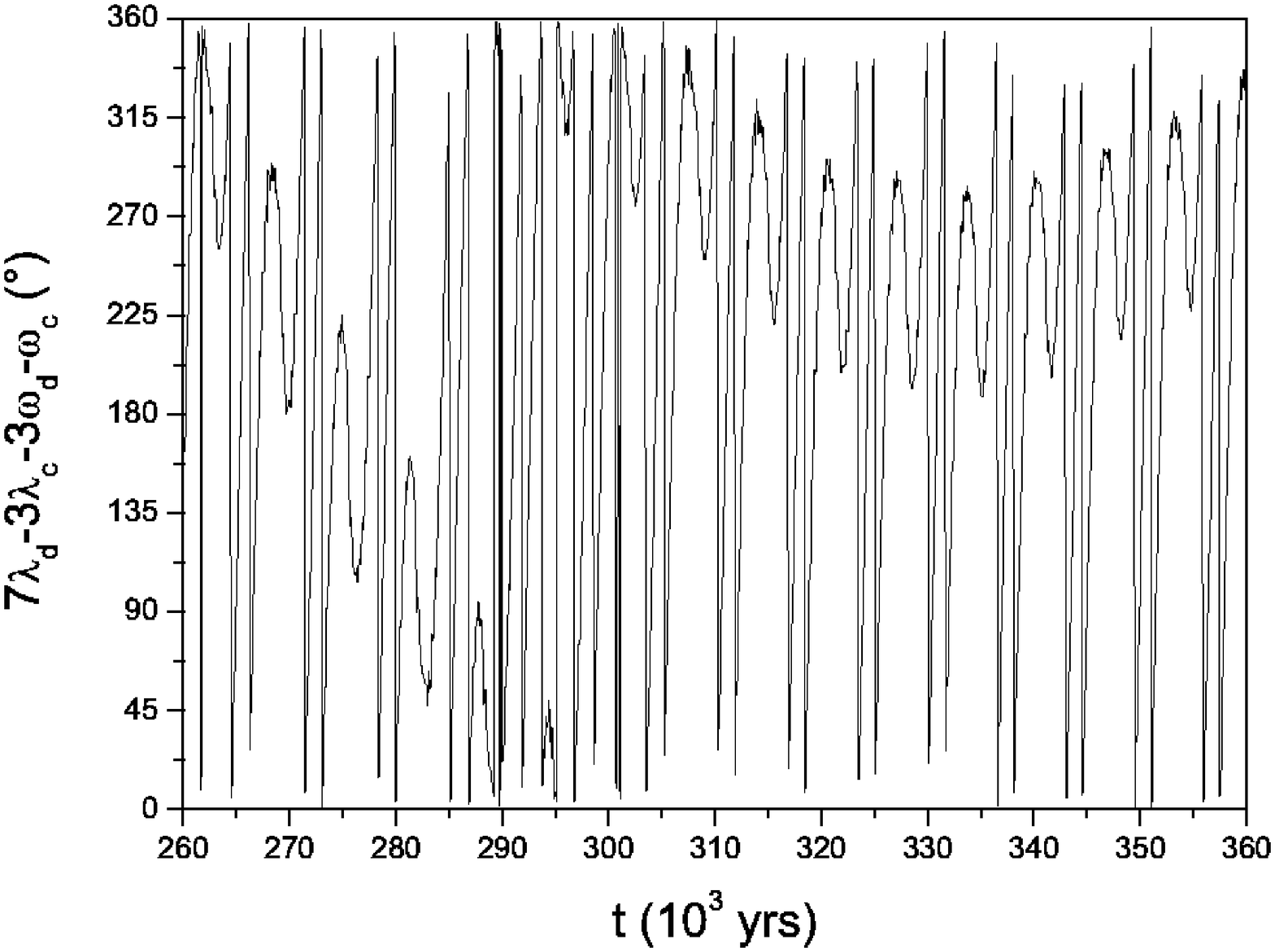}
\caption{Evolution of some orbital elements for the model nMMR 7:3. Left Panel: The solid line (black) indicates planet \textit{b}, the dashed line (red) denotes planet \textit{c} and the dotted line (blue) represents planet \textit{d}. $e_c$ moves in the interval 0.17-0.30 while $e_d$ in the range 0.15-0.30. This model is observed to be in SAR.}
\label{73}
\end{figure*}

\begin{figure*}
\includegraphics[width=0.5\textwidth]{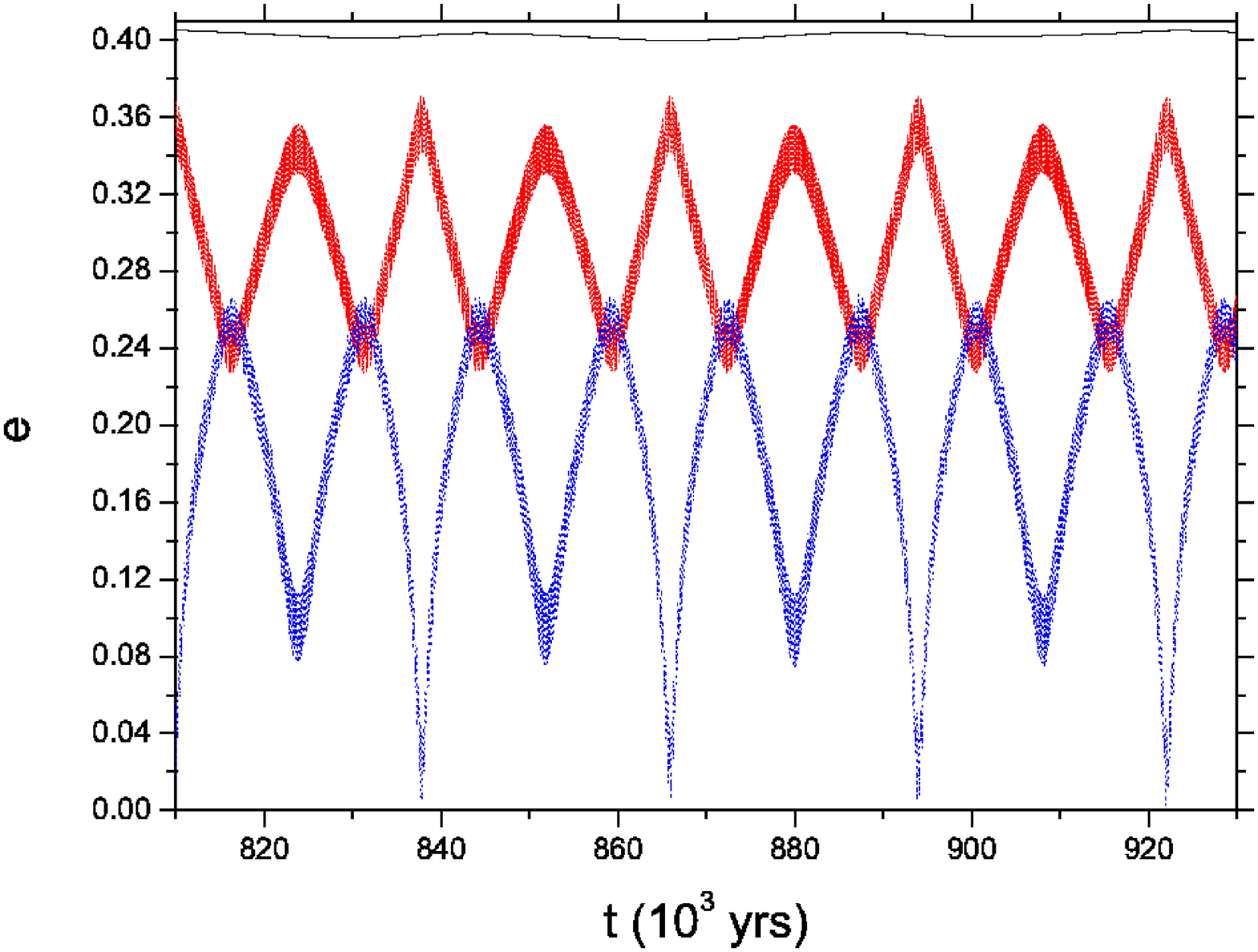}\includegraphics[width=0.5\textwidth]{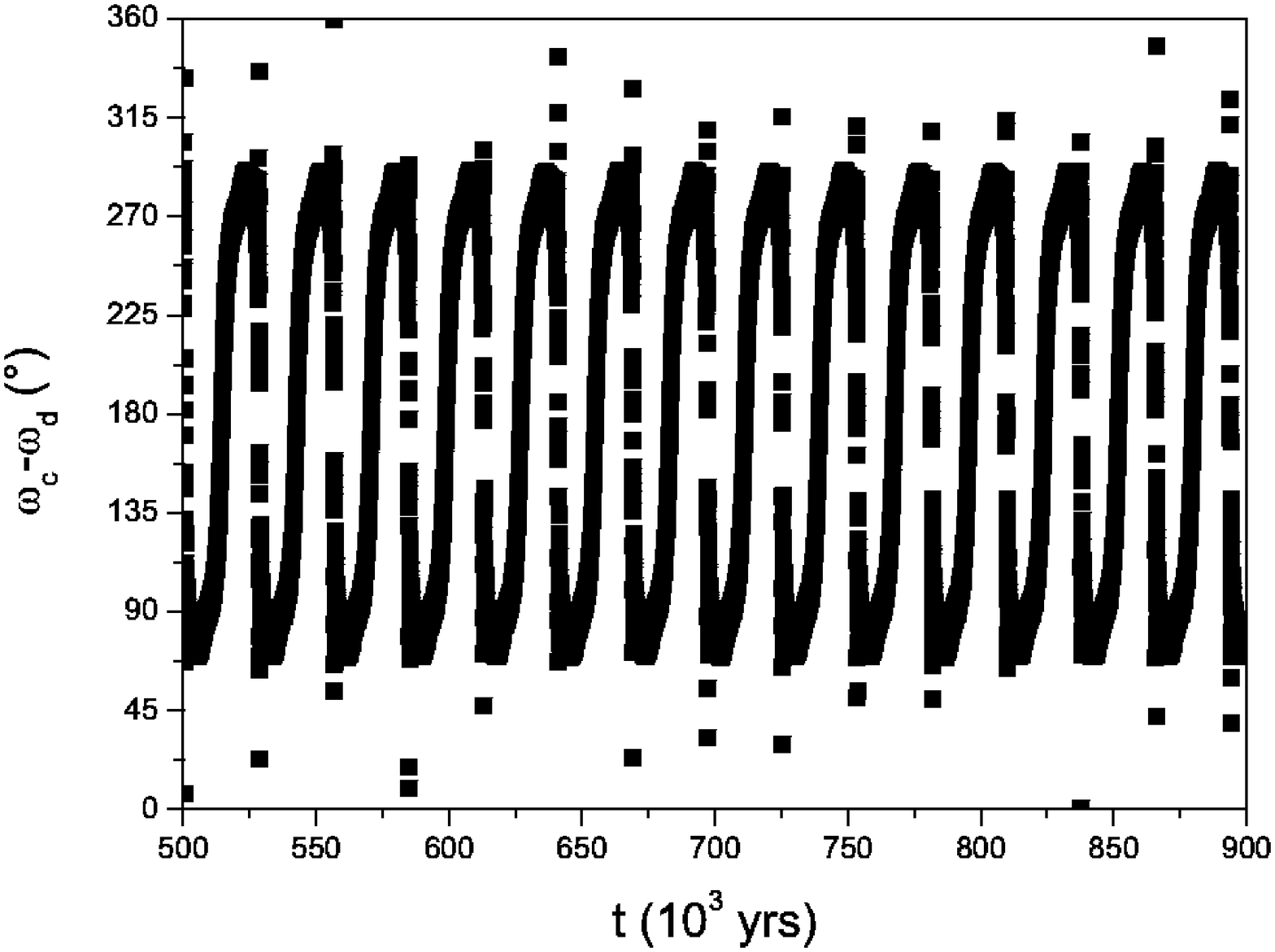}
\includegraphics[width=0.5\textwidth]{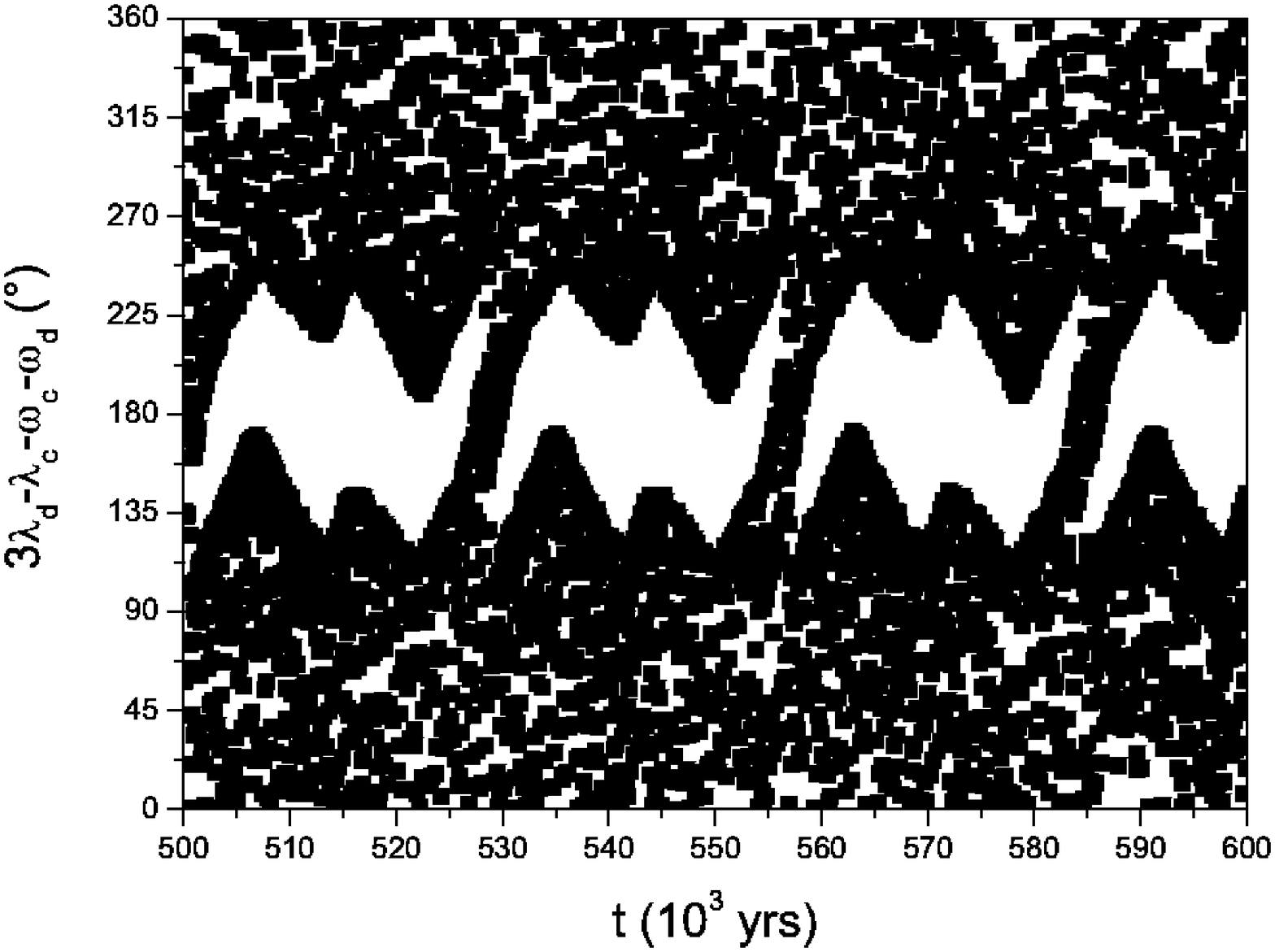}\includegraphics[width=0.5\textwidth]{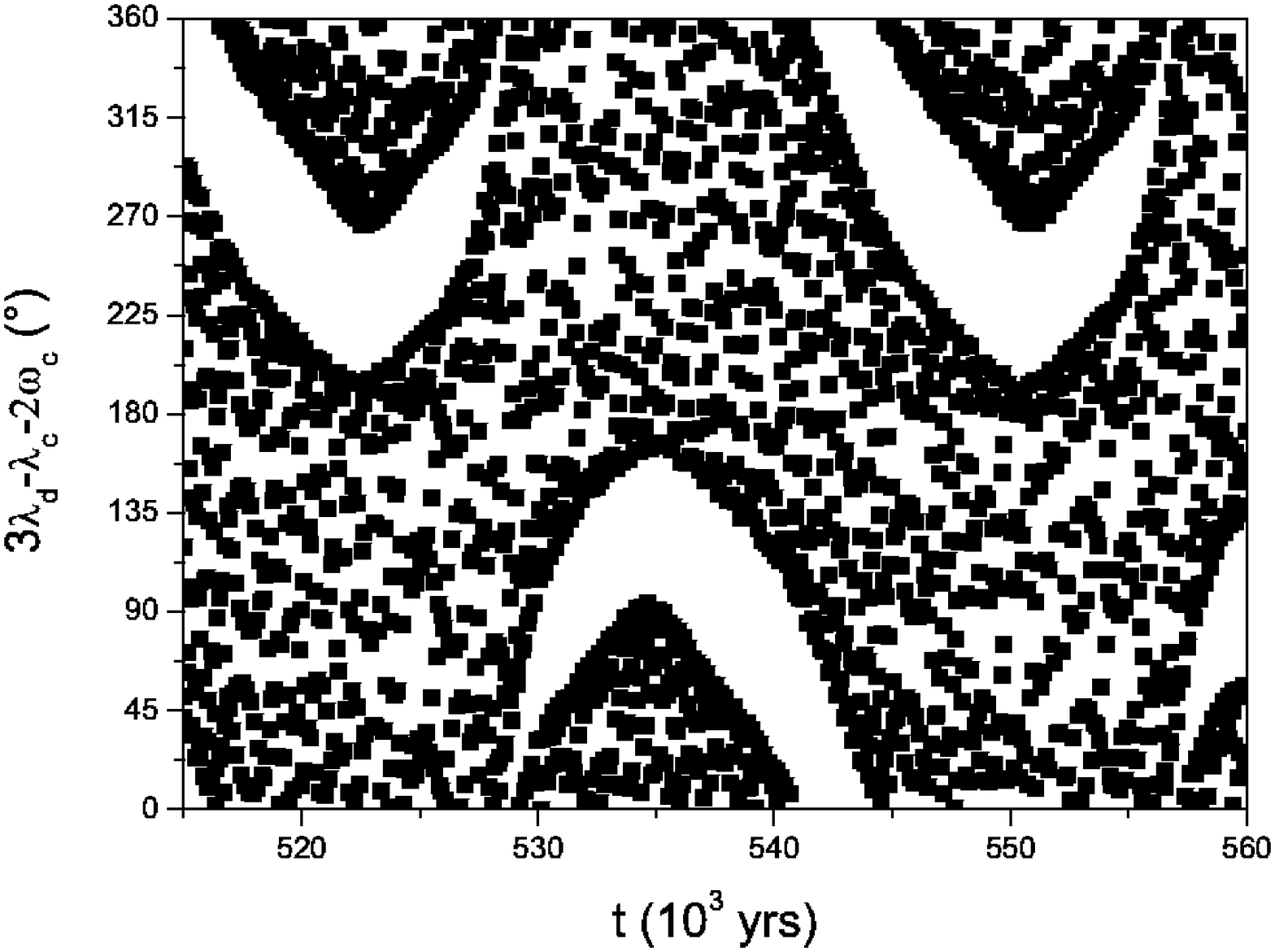}
\caption{Evolution of some orbital elements for the model nMMR 3:1. Top-left: The solid line (black) indicates planet \textit{b}, the dashed line (red) denotes planet \textit{c} and the dotted line (blue) represents planet \textit{d}. $e_c$ moves in the interval 0.23-0.37 while $e_d$ in the range 0.0-0.27. This model is in SAR.}
\label{31}
\end{figure*}

\begin{figure*}
\includegraphics[width=0.33\textwidth]{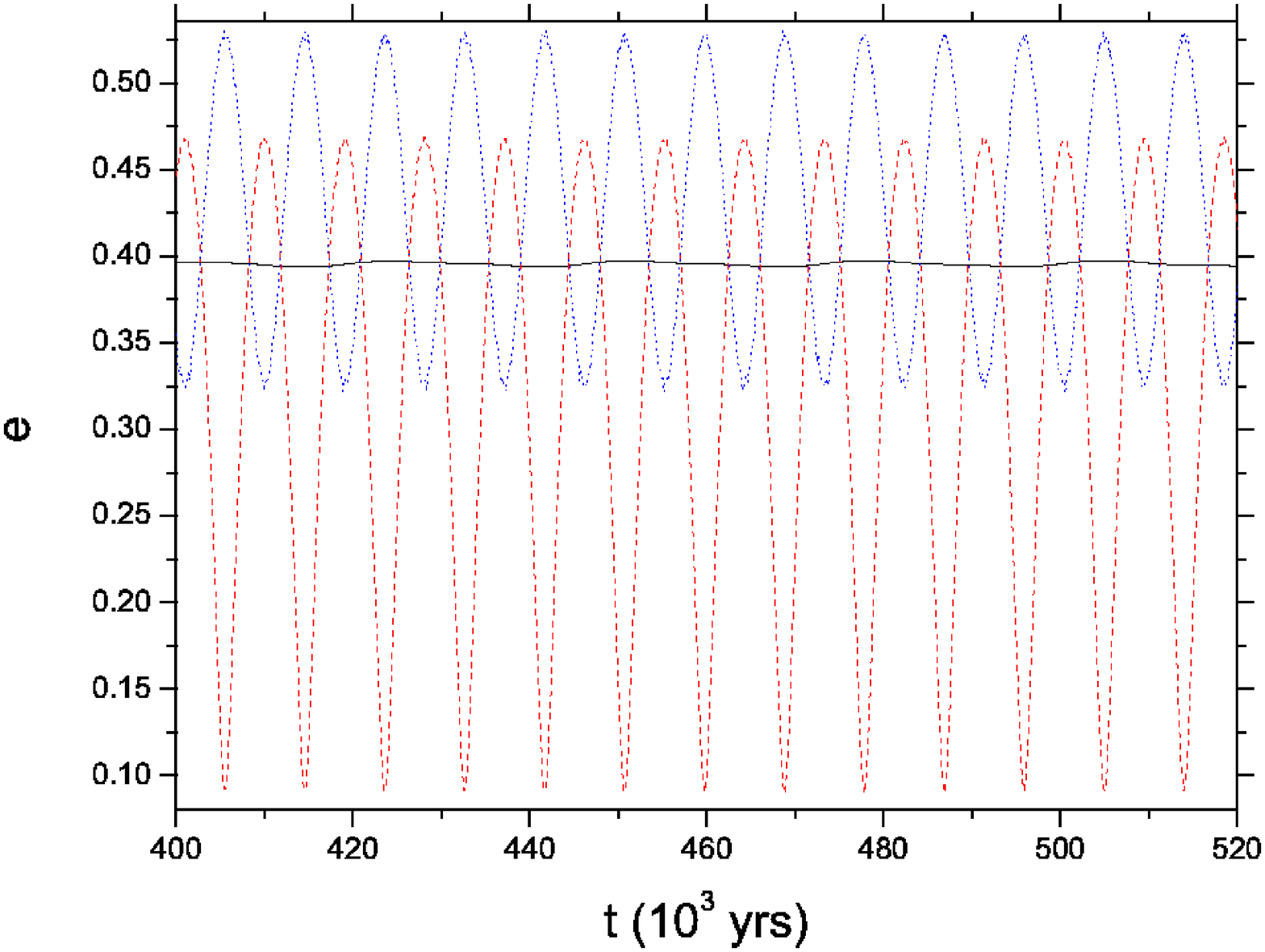}\includegraphics[width=0.33\textwidth]{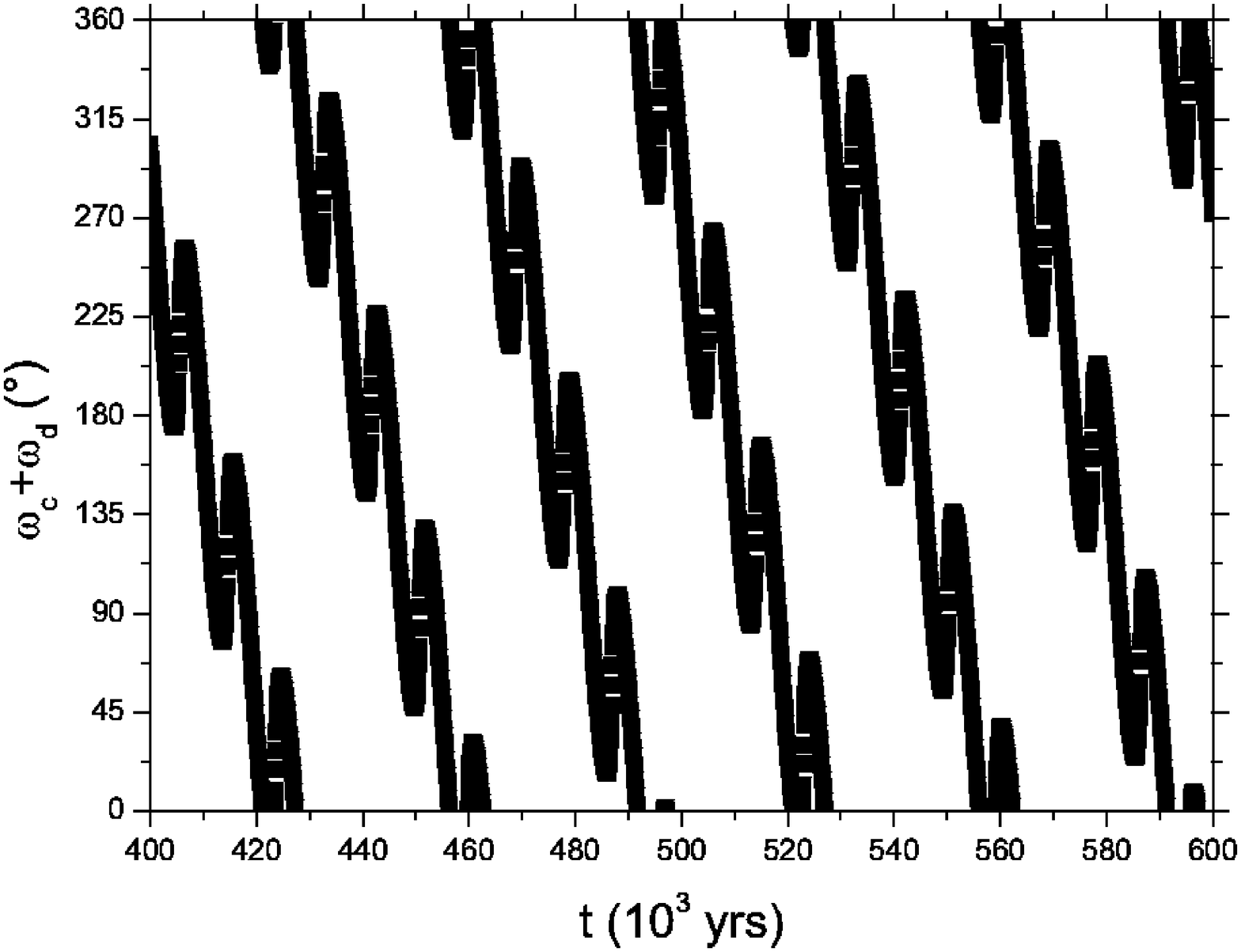}\includegraphics[width=0.33\textwidth]{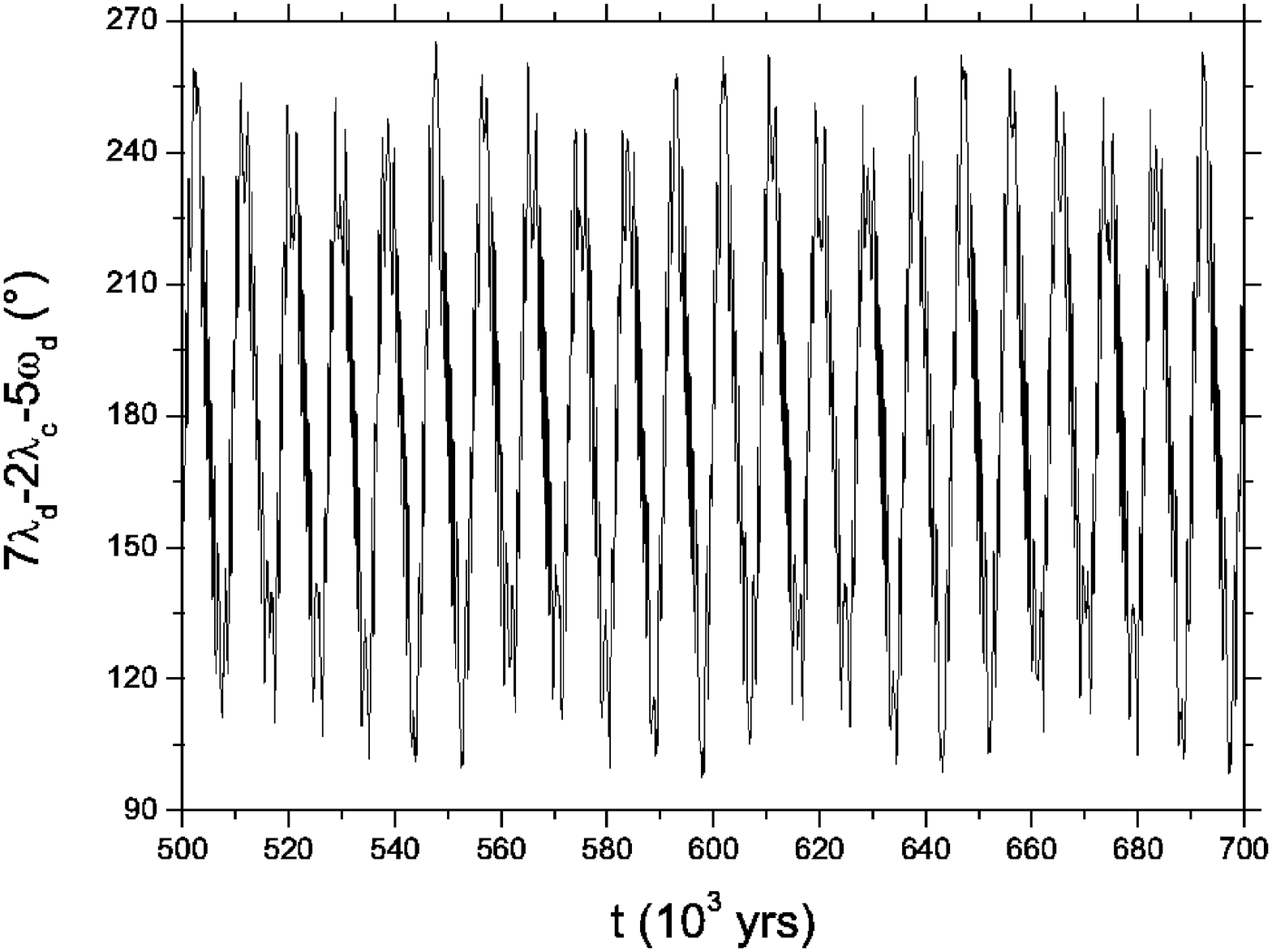}
\caption{Evolution of some orbital elements for the model nMMR 7:2. Left Panel: The solid line (black) indicates planet \textit{b}, the dashed line (red) denotes planet \textit{c} and the dotted line (blue) represents planet \textit{d}. $e_c$ moves in the interval 0.09-0.47 while $e_d$ in the range 0.32-0.53. The critical argument illustrated in the right panel librates around 180$^{\circ}$ with a semiamplitude of about 85$^{\circ}$.}
\label{72}
\end{figure*}

We compute the evolution of the orbital elements for fits corresponding to MMRs 12:5, 7:3, 3:1 and
7:2 which our simulations have demonstrated to preserve stability for at least 40 Myrs. The results
are illustrated in Figs. \ref{125}, \ref{73}, \ref{31} and \ref{72} and show the complexity of the possible dynamical
behaviours of the HD 181433 system that are consistent with the RV observations.

The case we show nMMR 12:5 has $\chi^2_{red} = 5.91$
and rms scatter of 1.46 m/s with $12n_d-5n_c \approx 1.8 ^{\circ}/yr$. This model is in SAR with $\omega_c-\omega_d$ librating around 0$^{\circ}$ with a semiamplitude of about 45$^{\circ}$. Moreover, we have found some critical arguments of the MMR to alternate librations with
circulations implying the resonance excites a chaotic configuration. The time evolution of the
critical angles $12\lambda_d-5\lambda_c-7\omega_c$ and $12\lambda_d-5\lambda_c-5\omega_d-2\omega_c$ are illustrated in Fig. \ref{125}.

The scenario nMMR 7:3 has $\chi^2_{red} = 5.89$ and rms scatter of 1.46 m/s with $7n_d-3n_c \approx 8.7 ^{\circ}/yr$. This model is observed to be in SAR with the critical angle $\omega_c-\omega_d$ librating around 0$^{\circ}$ with a semiamplitude of about 40$^{\circ}$ meaning that their periastrons are aligned. We have studied the resonant arguments of the 7:3 MMR and found that three angles i.e. $7\lambda_d-3\lambda_c-4\omega_d$, $7\lambda_d-3\lambda_c-3\omega_d-\omega_c$ and $7\lambda_d-3\lambda_c-2\omega_d-2\omega_c$, alternate librations with circulations. This indicates that the configuration is close to the resonance separatrices. Fig. \ref{73} illustrates how the critical angle $7\lambda_d-3\lambda_c-3\omega_d-\omega_c$ evolves with time.

The case near the low-order MMR 3:1 has $\chi^2_{red} = 6.03$ and rms scatter of 1.47 m/s with $3n_d-n_c \approx 0.6 ^{\circ}/yr$. This model is in SAR with $\omega_c-\omega_d$ librating around 180$^{\circ}$ with a semiamplitude of about 110$^{\circ}$ implying that this time the periastrons of planets \textit{c} and \textit{d} are anti-aligned. The data points that diverge from the periodic signal (Fig. \ref{31} top-right panel) represent the instants when $e_d$ gets close to be null and so the argument of pericentre, $\omega_d$, is not well defined. Furthermore, for the critical arguments of the MMR we find that librations alternate with circulations indicating a chaotic zone spanned by overlapping resonances. The time evolution of the critical angles $3\lambda_d-\lambda_c-\omega_d-\omega_c$ and $3\lambda_d-\lambda_c-2\omega_c$ are also illustrated in Fig. \ref{31}.

The scenario nMMR 7:2, illustrated in Fig. \ref{72}, has $\chi^2_{red} = 5.65$ and rms scatter of 1.45 m/s with $7n_d-2n_c \approx -1.7 ^{\circ}/yr$. This model is not seen to be in SAR. However, we find it to be locked in the MMR as the critical argument $7\lambda_d-2\lambda_c-5\omega_d$ librates around 180$^{\circ}$ with a semiamplitude of about 85$^{\circ}$. It is worth notice how this mechanism is capable of pumping $e_c$ from 0.09 to 0.47 in less than 5000 years. This configuration is located over the collision line, it is the resonance that prevents close encounters and provides long-term orbital stability to the system.

Considering all the the configurations studied, the behaviour of planet \textit{b} seems to be unrelated to the two giant companions as the amplitude of the eccentricity signal of \textit{b} appears to be unaffected even in the cases in which \textit{c} and \textit{d} are trapped in a resonance. Moreover, it seems unrealistic that planet \textit{b} can be involved in a p:q:r MMR with the outer two planets: \textit{b} is too far away from them and, in addition, it is mainly influenced by General Relativistic and tidal effects (a discussion of these mechanisms goes beyond the aim of this paper).

\begin{figure*}
\includegraphics[width=0.5\textwidth]{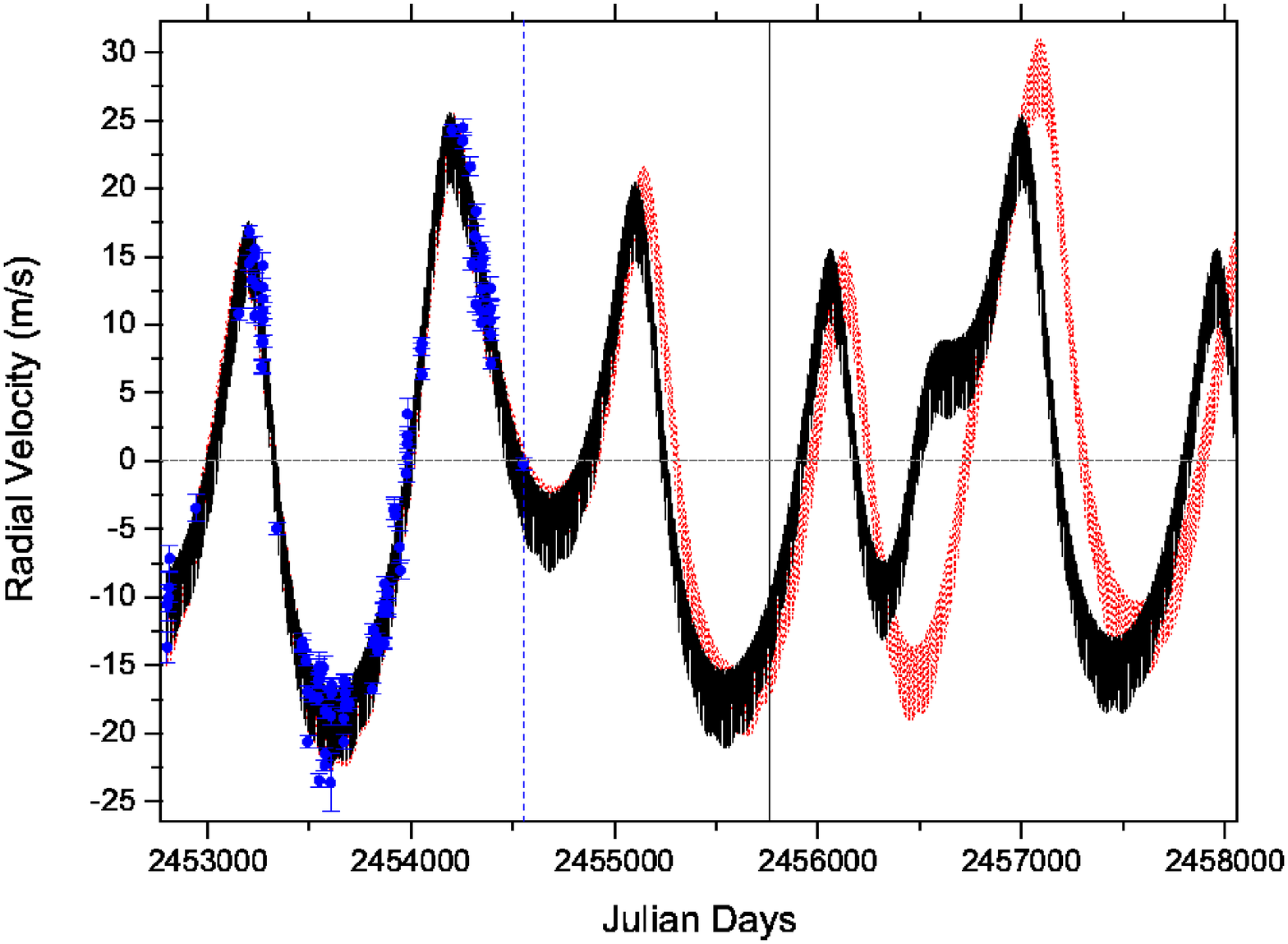}\includegraphics[width=0.5\textwidth]{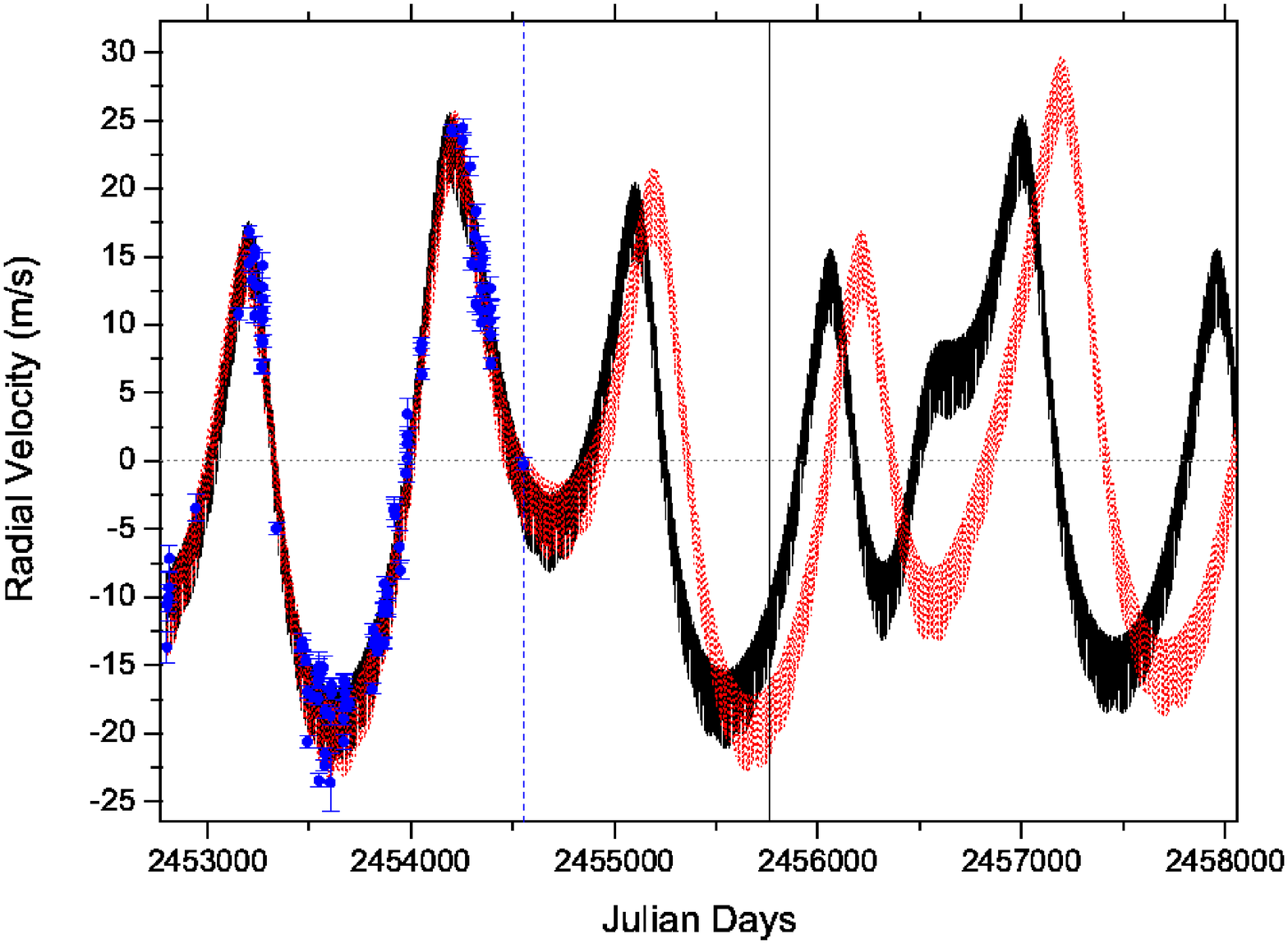}
\includegraphics[width=0.5\textwidth]{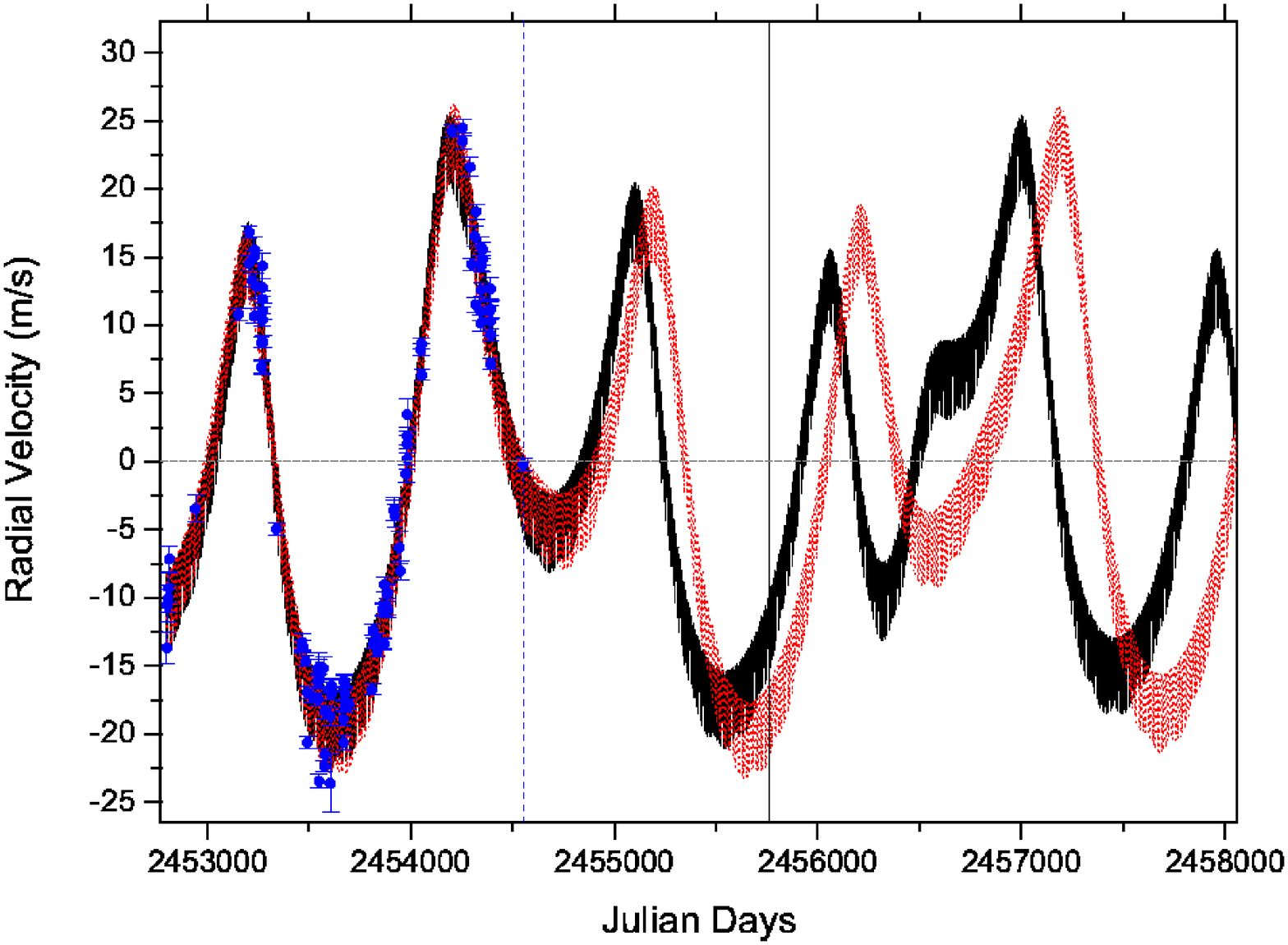}\includegraphics[width=0.5\textwidth]{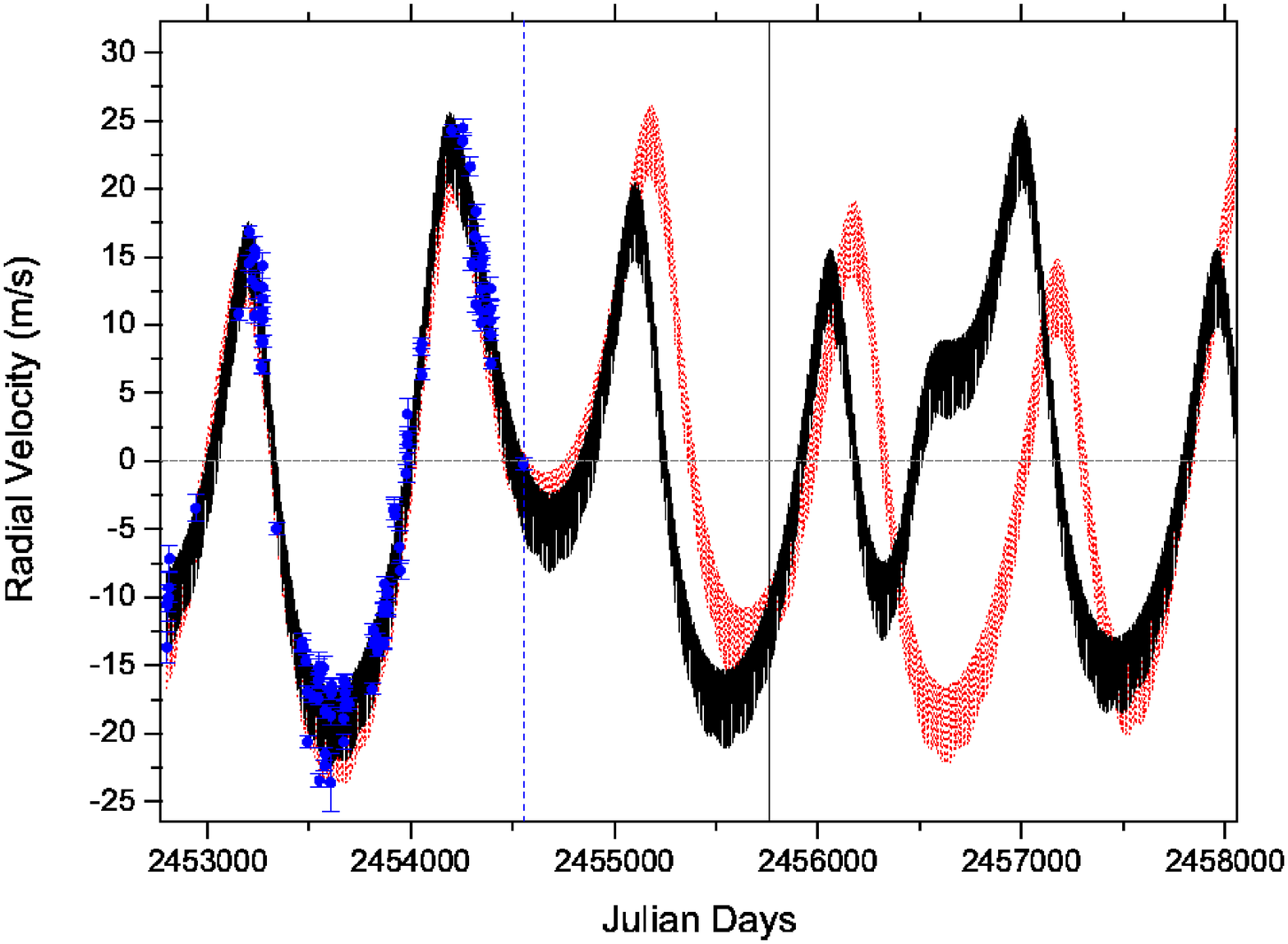}
\caption{Synthetic RV curves for HD 181433. In every panel, the straight line (black) indicates the Keplerian best-fit while the dotted line (red) represents (clockwise starting from the top-left) the stable best-fit, the configuration studied near the MMR 12:5, the model nMMR 7:2 and the scenario nMMR 7:3. Data points (blue) are plotted with error bars (and $V_{offset}$ is the one calculated for the stable best-fit). The dashed (blue) vertical line indicates when the last data point was taken. The straight (black) vertical line at JD 2,455,760 corresponds to the calendar date 2011 July 17 when the difference between most of the curves can be identified.}
\label{synth}
\end{figure*}

The synthetic RV curves for the Keplerian best-fit, stable best-fit and the models nMMRs 12:5, 7:3 and 7:2 are illustrated in Fig. \ref{synth}. The period through the year 2017 is covered. It is difficult to distinguish the curves from each other in the time range covered by the observations, but at the time of writing the difference can be spotted. However, the signal of the Keplerian best-fit will diverge evidently from the stable best-fit only in February 2013 ($\sim$ JD 2,456,340).

\section{DISCUSSION AND CONCLUSION}
\label{concl}
In this paper, our efforts have been focused on finding a plausible solution to the available RV data for the planetary system of HD 181433. In our investigation we have included an analysis of the long-term evolution of the system and the results support the thesis the dynamics is an important observable that has to be taken into account with the same priority as the RV observations.

The story with HD 181433 does not differ from the one of many other multi-planet systems found on the edge of long-term dynamical stability. The dynamical modelling of the RV with stability constraints offers precious information on the dynamical architecture of the putative planetary configurations. The stability criterion becomes a fundamental tool which provides limits to the physical and orbital elements of the planets when data do not cover completely the longest orbital period.

Our investigation leads to a Newtonian model for HD 181433 stable for at least 250 Myrs. The solution is compatible with what was found by \citet{b4}, but our analysis strongly diminish the uncertainity on the location of planet \textit{d} to the exiguous band where the 5:2 MMR is possible and stability is preserved. This seems the only plausible way to explain a very large eccentricity for the outermost planet, a quality which must be met in order to hold a good fit to the present data. In general, we can say that when an unstable high eccentric solution is found for a multiplanet system, the study of resonances may lead to the finding of a reasonable stable solution. By doing so, it is possible to constrain with high confidence the orbital period of the outermost (poorly sampled) planet well before sufficent data, covering several orbital periods, become available.

Apart the 5:2 MMR, the orbital evolution of the two giant companions is confined to a zone spanned by a number of other low-order two-body MMRs. We have studied different plausible stable configurations for the planetary system and, in particular, we have illustrated the behaviours caused by secular apsidal resonances and mean motion resonances. We have also found that at the time of writing with new data points it is definitely feasible to refine the circle of likely scenarios.

Furthermore, given the strong gravitational interactions between the two giant planets, a self-consistent N-body model for the RV data will help in estimating the inclination of the planetary orbits and of the physical masses (the RV method returns just the minimum masses for the planets). If it is not possible to use in synergy other methods e.g. astrometry, transits, it may be necessary to observe around 50 full orbits of planet  \textit{d} i.e. 300 years of RV data, to strongly constrain the orbital inclinations. As the large values already observed for the eccentricities may have been trigged by mechanisms which influence also the orbital inclinations (e.g. Libert \& Tsiganis 2009), a considerable value for their mutual inclination may be expected.

We can calculate the habitable zone (HZ) orbital distance $a_{hab}$, to be defined as the distance where a planet would receive the same insolation as the Earth: $a_{hab} = \sqrt{\frac{L_*}{L_{\odot}}}$ AU. For HD 181433, we find $a_{hab} = 0.55$ AU ($P_{hab} \sim$ 170 days) which is in the region between planets \textit{b} and \textit{c}. Our simulations, run for 40 Myrs, shows an Earth-size planet in the HZ (and in eccentric orbit) can retain stability indeed. The existence of a planet in this zone, not only fills an empty gap in the system, but would also offer a harbour for life. Therefore, this hypothetical planet becomes interesting for a double reason. Additional observations are required to investigate on the presence of further bodies and in particular on the existence of a terrestrial planet in the habitable zone of HD 181433.

The planetary system of HD 181433 offers a wide range of challenges that can go from understanding its real physical architecture to the study of potentially habitable worlds. Further observations can confirm the results illustrated in this paper, improve our understanding of the dynamical structure of this system and, in general, give additional insights into the study of the dynamics of planetary systems.

\section*{Acknowledgments}

The author acknowledges support of studentship from Queen Mary University of London and would like to thank Craig Agnor, Carl Murray, Stefano Meschiari and Richard Nelson for useful discussions and suggestions while conducting this research. We are also grateful to the anonymous referee for the valuable comments which improved the results presented in this paper.

\bsp

\label{lastpage}


\begin{thebibliography}{99}
\bibitem[\protect\citeauthoryear{Adams \& Laughlin}{2006}]{b1} F. Adams \& G. Laughlin. Effects of secular interactions in extrasolar planetary systems. ApJ, 649: 992, 2006
\bibitem[\protect\citeauthoryear{Anglada-Escud\'e et al.}{2010}]{b2} G. Anglada-Escud\'e, M. Lopez-Morales, J. Chambers. How eccentric orbital solutions can hide planetary systems in 2:1 resonant orbits. ApJ, 709: 168, 2010
\bibitem[\protect\citeauthoryear{Barnes \& Raymond}{2004}]{b3} R. Barnes \& S. Raymond. Predicting planets in known extrasolar planetary systems. I. ApJ, 617: 569, 2004
\bibitem[\protect\citeauthoryear{Bouchy et al.}{2009}]{b4} F. Bouchy et al. The HARPS search for southern extra-solar planets XVII. A\&A, 496: 527, 2009
\bibitem[\protect\citeauthoryear{Correia et al.}{2010}]{b5} A. Correia et al. The HARPS search for southern extra-solar planets XIX. A\&A, 511: A21, 2010
\bibitem[\protect\citeauthoryear{Go\'zdziewski et al.}{2003}]{b6} K. Go\'zdziewski, M. Konacki, A. Maciejewski. Where is the second planet in the HD 160691 planetary system? ApJ, 594: 1019, 2003
\bibitem[\protect\citeauthoryear{Go\'zdziewski et al.}{2006}]{b7} K. Go\'zdziewski, M. Konacki, A. Maciejewski. Orbital configurations and dynamical stability of multiplanet systems around sun-like stars HD 202206, 14 Herculis, HD 37124, and HD 108874. ApJ, 645: 688, 2006
\bibitem[\protect\citeauthoryear{Go\'zdziewski \& Migaszewski}{2009}]{b8} K. Go\'zdziewski \& C. Migaszewski. Is the HR 8799 extrasolar system destined for planetary scattering? MNRAS, 397: L16, 2009
\bibitem[\protect\citeauthoryear{Go\'zdziewski et al.}{2008}]{b9} K. Go\'zdziewski, C. Migaszewski, A. Musielinski. Stability constraints in modeling of multi-planet extrasolar systems. Proceedings IAU Symposium No. 249, 2008
\bibitem[\protect\citeauthoryear{Levison \& Duncan}{1994}]{b10} H. Levison \& M. Duncan. The long-term dynamical behavior of short-period comets. Icarus, 108: 18, 1994
\bibitem[\protect\citeauthoryear{Libert \& Tsiganis}{2009}]{b11} A.-S. Libert \& K. Tsiganis. Kozai resonance in extrasolar systems. A\&A, 493: 677, 2009
\bibitem[\protect\citeauthoryear{Marcy et al.}{2005}]{b12} G. Marcy et al. Five new extrasolar planets. ApJ, 619: 570, 2005
\bibitem[\protect\citeauthoryear{Mayor et al.}{2009a}]{b14} M. Mayor et al. The HARPS search for southern extra-solar planets XIII. A\&A, 493: 639, 2009a
\bibitem[\protect\citeauthoryear{Mayor et al.}{2009b}]{b15} M. Mayor et al. The HARPS search for southern extra-solar planets XVIII. A\&A, 507: 487, 2009b
\bibitem[\protect\citeauthoryear{Meschiari et al.}{2009}]{b16} S. Meschiari et al. Systemic: a testbed for characterizing the detection of extrasolar planets. I. PASP, 121: 1016, 2009
\bibitem[\protect\citeauthoryear{Meschiari et al.}{2011}]{b17} S. Meschiari et al. The Lick-Carnegie survey: four new exoplanet candidates. ApJ, 727: 117, 2011
\bibitem[\protect\citeauthoryear{Murray \& Holman}{2001}]{b18} N. Murray \& M. Holman. The role of chaotic resonances in the Solar System. Nature, 410: 773, 2001
\bibitem[\protect\citeauthoryear{Pepe et al.}{2003}]{b19} F. Pepe et al. From CORALIE to HARPS: towards 1 meter/sec RV precision. ASP Conf. Ser. 294: 39, 2003
\bibitem[\protect\citeauthoryear{Press et al.}{1992}]{b20} W. Press et al. Numerical recipes in C: the art of scientific computing (2nd ed.). Cambridge Univ. Press, 1992
\bibitem[\protect\citeauthoryear{Rivera et al.}{2005}]{b21} E. Rivera et al. A $\sim$7.5 $M_\oplus$ planet orbiting the nearby star, GJ 876. ApJ, 634: 625, 2005
\bibitem[\protect\citeauthoryear{Robutel \& Laskar}{2001}]{b22} P. Robutel \& J. Laskar. Frequency map and global dynamics in the Solar System I. Icarus, 152: 4, 2001
\bibitem[\protect\citeauthoryear{Sousa et al.}{2008}]{b23} S. Sousa et al. Spectroscopic parameters for 451 stars in the HARPS GTO planet search program. A\&A, 487: 373, 2008
\bibitem[\protect\citeauthoryear{Veras \& Ford}{2010}]{b24} D. Veras \& E. Ford. Secular orbital dynamics of hierarchical two planet systems. ApJ, 715: 803, 2010
\bibitem[\protect\citeauthoryear{Vogt et al.}{2010}]{b25} S. Vogt et al. A super-Earth and two Neptunes orbiting the nearby Sun-like star 61 Virginis. ApJ, 708: 1366, 2010
\bibitem[\protect\citeauthoryear{Wright et al.}{2009}]{b26} J. Wright et al. Ten new and updated multiplanet systems and a survey of exoplanetary systems. ApJ, 693: 1084, 2009.
\end{thebibliography}
\end{document}